\documentclass[sigconf]{acmart}

\usepackage{booktabs} % For formal tables
\usepackage{graphicx,subfigure}
\usepackage{caption}
\usepackage{color}
\usepackage{tabularx}
\usepackage{pdfpages}

\AtBeginDocument{%
  \providecommand\BibTeX{{%
    \normalfont B\kern-0.5em{\scshape i\kern-0.25em b}\kern-0.8em\TeX}}}

\copyrightyear{2021}
\acmYear{2021}
\setcopyright{rightsretained}
\acmConference[CHI '21]{CHI Conference on Human Factors in Computing Systems}{May 8--13, 2021}{Yokohama, Japan}
\acmBooktitle{CHI Conference on Human Factors in Computing Systems (CHI '21), May 8--13, 2021, Yokohama, Japan}\acmDOI{10.1145/3411764.3445200}
\acmISBN{978-1-4503-8096-6/21/05}

\makeatletter
\let\@authorsaddresses\@empty
\makeatother

\begin{document}

\title{Significant Otter: Understanding the Role of Biosignals in Communication}

\settopmatter{authorsperrow=4}
\author{Fannie Liu}
\affiliation{
  \institution{Snap Inc.}
  \city{New York}
  \state{NY}
  \country{USA}
}
\additionalaffiliation{Carnegie Mellon University}
\email{fannie@snap.com}

\author{Chunjong Park}
\affiliation{
  \institution{University of Washington}
  \city{Seattle}
  \state{WA}
  \country{USA}
}
\email{cjparkuw@cs.washington.edu}

\author{Yu Jiang Tham}
\affiliation{
  \institution{Snap Inc.}
  \city{Seattle}
  \state{WA}
  \country{USA}
}
\email{yujiang@snap.com}

\author{Tsung-Yu Tsai}
\affiliation{
  \institution{Snap Inc.}
  \city{Seattle}
  \state{WA}
  \country{USA}
}
\email{ttsai@snap.com}

\author{Laura Dabbish}
\affiliation{
    \institution{Carnegie Mellon University}
    \city{Pittsburgh}
    \state{PA}
    \country{USA}
}
\email{dabbish@cs.cmu.edu}

\author{Geoff Kaufman}
\affiliation{
    \institution{Carnegie Mellon University}
    \city{Pittsburgh}
    \state{PA}
    \country{USA}
}
\email{gfk@cs.cmu.edu}

\author{Andr\'es Monroy-Hern\'andez}
\affiliation{
  \institution{Snap Inc.}
  \city{Seattle}
  \state{WA}
  \country{USA}
}
\email{amh@snap.com}

\renewcommand{\shorttitle}{Significant Otter: Understanding the Role of Biosignals in Communication}
\renewcommand{\shortauthors}{Liu et al.}

\begin{abstract}
With the growing ubiquity of wearable devices, sensed physiological responses provide new means to connect with others. While recent research demonstrates the expressive potential for biosignals, the value of sharing these personal data remains unclear. To understand their role in communication, we created Significant Otter, an Apple Watch/iPhone app that enables romantic partners to share and respond to each other’s biosignals in the form of animated otter avatars. In a one-month study with 20 couples, participants used Significant Otter with biosignals sensing OFF and ON. We found that while sensing OFF enabled couples to keep in touch, sensing ON enabled easier and more authentic communication that fostered social connection. However, the addition of biosignals introduced concerns about autonomy and agency over the messages they sent. We discuss design implications and future directions for communication systems that recommend messages based on biosignals.
\end{abstract}


\maketitle

\begin{figure}[t!]
\centering
  \includegraphics[width=.85\columnwidth]{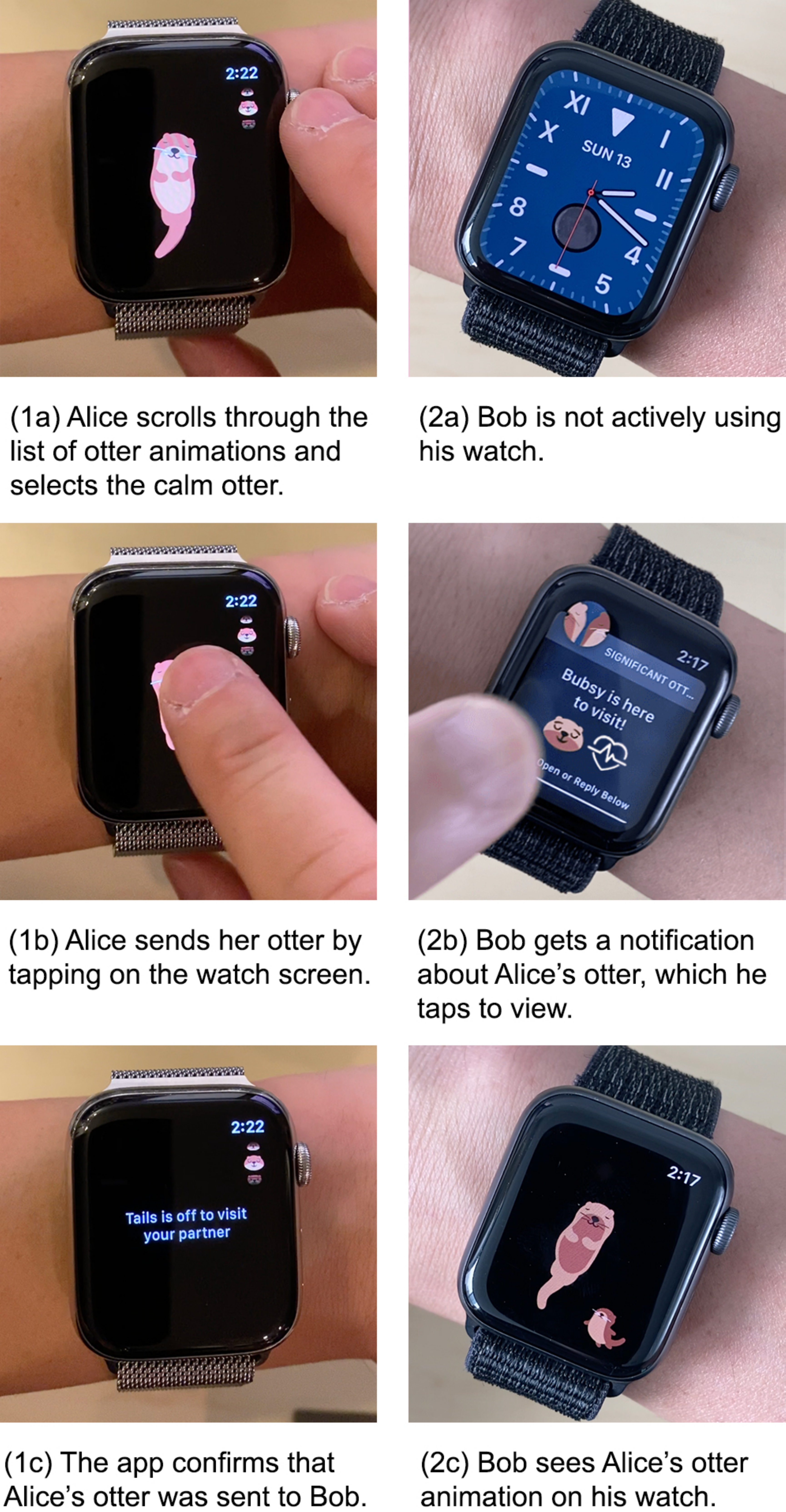}
  \caption{A hypothetical couple using Significant Otter: Alice (left) sends her state otter to Bob (right) on her Apple Watch.}~\label{fig:send_receive}
 \Description[Screenshots of the user flow for sending an otter state]{Screenshots of the user flow for sending an otter state on the Apple Watch. A hypothetical user Alice scrolls through her state otter list and taps on her otter. Tapping on it animates the otter off screen to visit Bob, who gets a notification from the app. Bob opens the notification and sees an icon of Alice's otter, which he taps to view the full animation of the otter.}
\end{figure}

\section{Introduction}

Today, we rely heavily on digital technology to connect with others. Furthermore, with the global COVID-19 pandemic diminishing in-person social contact, technology-mediated communication is more prominent than ever before~\cite{nytimes_covid}. However, digital communication is well-known to be challenging due to limited access to important nonverbal cues, such as our body movements and facial expressions~\cite{tanis2003social,kiesler1984social,walther2011theories}.

An emerging area of research in HCI has explored a novel social cue for improving the way we interact over technology: our biosignals. Biosignals, such as heart rate and skin conductance, are well known to change according to our physical and emotional responses, and can be revealed in everyday interactions using wearable sensor technologies. For example, applications like Pulsoid or Onbeat\footnote{\url{https://pulsoid.net/} and \url{http://onbeat.fit/}} explore this possibility through livestreams of heart rate during gameplay or exercise. Researchers have shown that \textit{expressive biosignals}, or biosignals displayed as a social cue, have the potential to facilitate communication as a means to recognize and express our emotions and physical being~\cite{liu2019animo,liu2017supporting,howell2016biosignals,semertzidis2020neo,liu2017expressive,liu2019empathy,janssen2010intimate,min2014biosignal,howell2019bench,slovak2012understanding}. However, researchers have not yet described the \textit{role} that biosignals play in communication. Biosignals are personal and private data that require careful design and consideration~\cite{liu2017supporting,janssen2010intimate,howell2018emotional,howell2018tensions}. In particular, as cues that are sensed and recommended by systems, they present a new form of AI-mediated communication that could shape our interactions in unintended ways~\cite{jakesch2019ai}. Thus, it is crucial that we understand the value and consequences of integrating them into our existing means of communicating.

In the present work, we expand on expressive biosignals literature by demonstrating the effects of shifting from communication \textit{without} biosignals to communication \textit{with} biosignals. We designed, developed, and deployed Significant Otter, an Apple Watch and iPhone app that enables romantic couples to send heart rate-driven otter animations as messages to each other. By setting adaptive thresholds for each person based on their past heart rate and motion data, Significant Otter intelligently suggests animations that match their current emotional and physical state. To explore the design of expressive biosignals as AI-mediated communication, we incorporate AI-recommended \textit{sets} of shareable sensed states. In a one-month within-subjects field study, we investigate how couples' behaviors and perceptions are affected when shifting from a sensing OFF version of the app, with no biosignals sensed, to a sensing ON version, with biosignals sensed. We present qualitative results from interviews during the study and discuss opportunities and challenges for biosignals in communication.

The core contributions of this work are: (1) Significant Otter\footnote{Significant Otter is publicly available on the App Store at the following link: \url{https://apps.apple.com/us/app/significant-otter-couples-app/id1450105275}}, a novel smartwatch and phone app that promotes communication and connection between romantic partners through animated avatars recommended based on heart rate; (2) an empirical study with 20 couples who used Significant Otter with sensing OFF and ON that demonstrates the value of biosignals as a lightweight and authentic social cue; (3) design implications and future directions for expressive biosignals research, including suggestions for integration into social platforms as a form of AI-mediated communication.

\section{Background}
\subsection{Research Context: Technology for Romantic Couples}
For the purposes of this study, we focus on communication between romantic couples. Given the intimate nature of physiological data~\cite{janssen2010intimate}, people feel most comfortable sharing them with close others~\cite{liu2017supporting}, who may also be the most interested and equipped to understand them as limited contextual cues. For instance, couples can interpret work breaks or distance from home based how many steps their partner has made~\cite{griggio2019augmenting}. In lightweight communication, defined by quick exchanges~\cite{cowan2011lightweight} through minimal interaction or content generation, even minimal messages between close partners can convey meaning like ``thinking of you''~\cite{kaye2006just,counts2004supporting}. Thus, we target the closest partners: significant others.

A breadth of HCI research has explored technologies that can support significant others, including those that integrate biosignals. In their review on technology-mediated intimacy, Hassenzahl and colleagues described different strategies for supporting important aspects of intimacy~\cite{hassenzahl2012all}. For example, physicalness represents the physical aspect of intimacy, and has been simulated through mediated touch~\cite{haans2006mediated} and gestures~\cite{eichhorn2008stroking,mueller2005hug}, as well as feeling someone else's heartbeat~\cite{werner2008united}. Expressivity describes expressing feelings through a language unique to the couple, such as mutual affection through ``on-off'' signals~\cite{kaye2005communicating} or couple-specific symbols~\cite{kowalski2013cubble}. Awareness of one's partner has been explored in systems that display a partner's presence, activities, and mood through availability~\cite{dey2006awareness,cho2020share} or sensed contextual information like location~\cite{wiese2011you,bales2011couplevibe}, motion~\cite{bentley2007sharing,wilson2003narrator}, and heart rate~\cite{hassib2017heartchat}, or a combination of these data~\cite{griggio2019augmenting}.

With the integration of biosignals, Significant Otter similarly incorporates physicalness, expressivity, and awareness to support intimate communication. Significant Otter can simulate physicalness through shared heart rate representing the body's physical state. It can support expressivity by providing an emotional language for couples through otter animations embedded with heart rate, which communication partners can use to create emotional meaning together~\cite{liu2017supporting}. Finally, the app's heart rate animations can enable awareness by providing contextual cues that display presence, activities, and mood~\cite{liu2019animo,hassib2017heartchat}. We describe the full Significant Otter system in Section~\ref{sec:system}.

\subsection{AI-Mediated Communication} Hancock and colleagues define AI-mediated communication as ``mediated communication
between people in which a computational agent operates on behalf of a communicator by modifying, augmenting, or generating messages to accomplish communication or interpersonal goals.'' They suggest that AI-mediated communication systems may have important effects on interpersonal dynamics, including self-presentation and disclosure, and subsequently, meaningful and intimate relationships~\cite{jakesch2019ai}. As a relatively new area of research, much of the AI-mediated communication work has focused on text, such as AI-recommended wording in emails~\cite{boomerang} and AI-generated profiles on sites like AirBnB~\cite{jakesch2019ai}. The present work expands on this research by exploring AI-mediated communication through expressive biosignals.

Expressive biosignal systems recommend a user's current state as part of interpersonal communication. The recommendation can be used to augment communication, such as by providing emotional context for text messages~\cite{liu2017supporting,hassib2017heartchat} and joint activities in mixed reality~\cite{semertzidis2020neo}, or to generate new messages in communication, such as emoji-like animations~\cite{liu2019animo}. Like other forms of AI-mediated communication, AI-recommended states through biosignals could impact key aspects of communication. For example, in their deployment of Ripple, a shirt that displayed a wearer's skin conductance, Howell and colleagues found that people granted the system high degrees of authority over their feelings and, therefore, the feelings they conveyed to others~\cite{howell2018tensions}. On the other hand, Liu and colleagues showed that some people may strongly disagree with an AI-recommended state, and subsequently fail to use the system to communicate meaningfully with others~\cite{liu2019animo}. To address these issues, we explore the design of an expressive biosignal system with a lower level of \textit{autonomy}, i.e., the degree of control it has over messages~\cite{jakesch2019ai}. Specifically, we explore communication in which people choose between shareable states suggested by the AI, rather than the AI providing only one possible state. To understand the effects of biosignals-based recommendations, we compare people's perceptions of their communication when they can share from a set of \textit{random} versus \textit{sensed} states, described in more detail in Section~\ref{sec:system}.

\subsection{Effects of Biosignals on Communication and Connection}
Existing literature on expressive biosignals have primarily explored their potential for supporting how we communicate and connect with each other. Following the interaction model of communication~\cite{west2018introducing}, these works suggest that biosignals can support the key stages of communication: sending a message, receiving and understanding that message, and responding to it with feedback. By sharing their biosignals as a message, a sender can express both emotions and daily activities, such as texting one's heart rate to convey feeling down or taking a walk~\cite{liu2017supporting}. Upon viewing the sender's biosignals, a receiver can become aware of the sender's state. Hassib and colleagues showed that when accessing someone's heart rate on a mobile messaging app, people can recognize when that person is angry or on their way home~\cite{hassib2017heartchat}. A recent controlled study also showed that biosignals increase emotional perspective-taking, or imagining someone else's emotions, in the context of a narrative story~\cite{liu2019empathy}. Receivers may subsequently respond with feedback based on their understanding of the sender's emotions. In prior studies where people shared their biosignals in conversation or sporadically during the day, receivers often acknowledged, provided support for, or discussed the meaning of the sender's biosignals ~\cite{liu2019animo,howell2016biosignals,howell2018tensions}.

Expressive biosignals may also impact social connection, or ``a person’s subjective sense of having close and positively experienced relationships with others in the social world''~\cite{seppala2013social}. According to Slov{\'a}k and colleagues, expressive biosignals may promote connectedness between people in two ways. First, they suggest that expressive biosignals are a form of emotional self-disclosure, as they can represent our internal emotional reactions during personal experiences~\cite{slovak2012understanding}. Self-disclosure is crucial for people to connect with each other, where it can improve the quality of interactions and closeness in relationships~\cite{aron1992inclusion,laurenceau1998intimacy}. Second, Slov{\'a}k and colleagues suggest that biosignals indicate a person's physical being as a representation of the daily physiological workings of our heart and other organs, thereby creating feelings of presence~\cite{slovak2012understanding}, which can lead to feelings of connectedness~\cite{ijsselsteijn2003staying}. 
For instance, Howell and colleagues showed that listening to someone's heartbeat on a bench can elicit a sense of being alive and connected to another living person~\cite{howell2019bench}. Moreover, Liu and colleagues suggest that remote interactants can feel present with each other when sharing their biosignals over smartwatches~\cite{liu2019animo}.

Although these prior works suggest the potential for biosignals to support communication and connection, they have not illustrated the \textit{value} of expressive biosignals. In particular, their emotionally expressive ability may already be achieved verbally and nonverbally through emojis and stickers~\cite{tang2018emoticon,lo2008nonverbal}. Since expressive biosignals can elicit concerns around privacy~\cite{liu2017supporting,slovak2012understanding}, cognitive load~\cite{liu2017expressive}, and accuracy~\cite{liu2019animo,howell2018tensions,merrill2019sensing}, it is crucial that we understand the value they add to existing modes of expression. Research suggests several possibilities for expressive biosignals to improve how we interact today, drawing from their potential to validate feelings as ``objective'' cues~\cite{liu2017supporting,semertzidis2020neo,howell2018tensions}. In the different stages of communication, biosignals sent as a message could be a more vivid way to express ourselves emotionally, understand those expressions directly from the body, and subsequently provide improved feedback. As vivid emotional expressions from our bodies, they may be perceived as more authentic and intimate disclosures of our internal experiences, leading to greater feelings of connection. We explore these possibilities by comparing communication with and without biosignals.

To our knowledge, only a few studies have compared the presence and absence of biosignals in social contexts~\cite{liu2019empathy,janssen2010intimate,curran2019empathy,merrill2017trust}. However, these works did not test in real-world dyadic communication, instead focusing on perceptions of a target other in controlled laboratory settings. We address this gap through a field study with couples who used two versions of Significant Otter, an expressive biosignals app. Specifically, we investigate communication and social connection between couples who shift from a sensing OFF version, with no biosignals, to sensing ON, with biosignals.

\begin{quote}
\textbf{RQ1:} How does shifting from sensing OFF to sensing ON affect the stages of communication (sending, understanding, responding) between couples?
\end{quote}
\begin{quote}
\textbf{RQ2:} How does shifting from sensing OFF to sensing ON affect social connection between couples?
\end{quote}

We designed Significant Otter based on prior expressive biosignals systems tested in everyday contexts~\cite{liu2017supporting,liu2019animo}. Like these works, we focus on heart rate due to its wide availability on consumer-grade wearables compared to other biosignals, representing the data as an animated avatar. However, unlike prior systems, Significant Otter recommends a \textit{set} of heart rate-driven avatars, rather than a single number or avatar, in order to further explore biosignals in AI-mediated communication. We detail our system in the following section.

\section{Significant Otter System}\label{sec:system}
Significant Otter is an Apple Watch and iPhone app that enables two people to send animated otter characters to each other based on their biosignals. Each person has an animated otter that reflects their inner state, which they can send to their partner. We designed Significant Otter to provide a playful way for couples to communicate. The app has been publicly available since November 2019, and over 59 thousand people have installed it as of January 2021.

To investigate our research questions, we created two study versions\footnote{Study versions of Significant Otter were \textit{not} publicly available. Public users were not included in our study.} of the Significant Otter app: sensing ON and sensing OFF, where biosignals are either sensed or not sensed, respectively. With sensing ON, people can send otter animations from a list of \textit{sensed} states, suggested based on their biosignals. With sensing OFF, people can send otter animations from a list of \textit{random} states, randomly selected by the system. For both versions, the app prompts people to name their otter and pair with the partner. Sensing ON requires people to accept HealthKit\footnote{HealthKit is Apple's framework for health and fitness data, such as from recorded vital signs and workouts.} and Motion \& Fitness\footnote{This refers to Apple's Core Motion framework for motion data, such as accelerometer and gyroscope data.} permissions to access sensed heart rate and activity data from the watch. People can then view their otter on their watch or phone, and scroll through the list of animated states to send one to their partner. We developed Significant Otter as a watch-first app, since smartwatches can be an unobtrusive and lightweight platform for communicating biosignals~\cite{liu2019animo}, but we included a phone version due to Apple's watch app requirements at the time.

\subsection{Pilot Studies}
We ran two pilot studies as initial tests for Significant Otter. The first pilot tested people's understanding of the sensing ON version, to ensure that people would recognize that their heart rate is sensed and tied to the animations. The pilot included seven couples (employees of a technology company and their significant others) who used the app freely for one week and were asked about their usage and perceptions of the app. Based on their responses, we iterated on the app to improve its usability. For instance, we determined the final list of animations based on the pilot results, which suggested that we should include states that people typically relate to heart rate (e.g., exercise), as well as limit the number of available animations for usability. The second pilot tested transitioning from sensing OFF to sensing ON. We ran this pilot for three weeks with three couples (employees of the same technology company and their significant others). The results confirmed that people viewed sensing ON as a feature update that included biosignals, and provided initial insights that informed the development of the study materials. We also ran this pilot during the early stages of the COVID-19 stay-at-home orders for many states in the United States. Stay-at-home orders required that people stay in their residences except for essential trips, such as for daily food and supplies or if they were essential workers (e.g., life-sustaining occupations, including employees in healthcare, food retail, and public transportation). The pilot informed ways to address possible COVID-19 circumstances that would affect the study. The final versions of the study app and study design are described in the following sections.

\begin{figure}[b!]
\centering
  \includegraphics[width=\columnwidth]{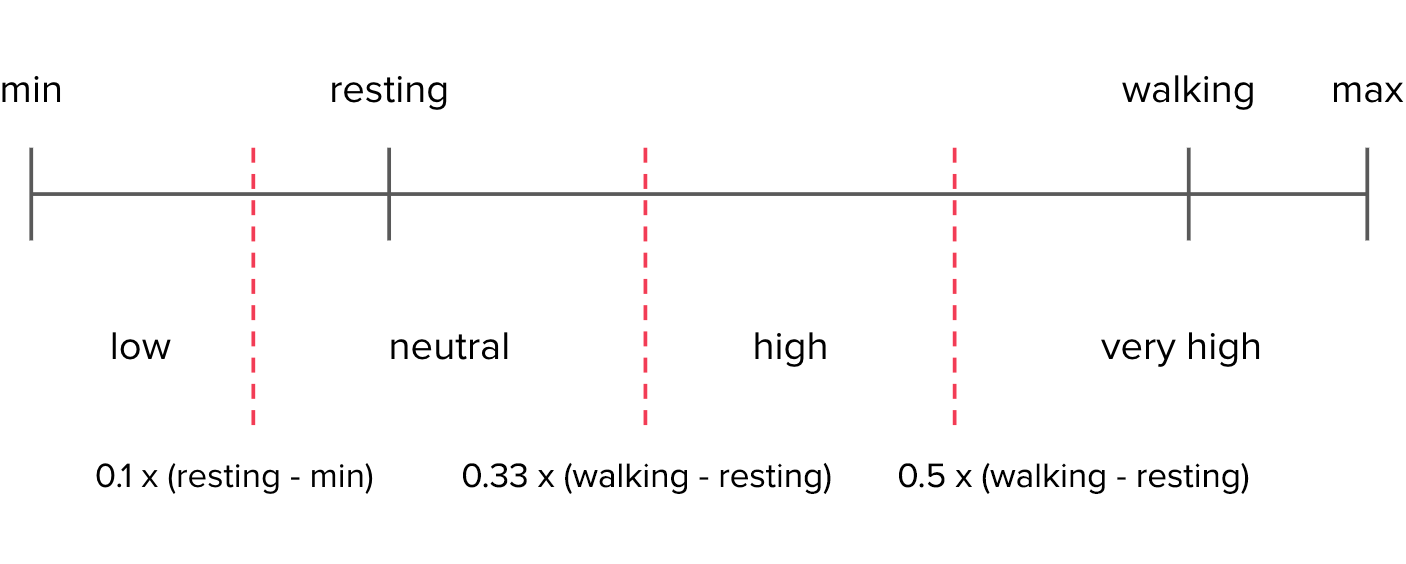}
  \caption{Heart rate sensing based on people's historical data from Apple HealthKit, with thresholds determined through empirical testing. The thresholds are relative to being stationary, according to prior work in emotion detection using physiological data~\cite{egger2019emotion}.}
  ~\label{fig:hrthresholds}
\Description[Thresholds for low to very high heart rate ranges based]{Thresholds for low, neutral, high, and very high heart rate ranges based on min, resting, walking, and max heart rate provided by Apple HealthKit.}
\end{figure}

\subsection{Otter Animations}
People can send two types of animations to their partner: \textbf{states} and \textbf{reacts}. People can send states to initiate communication, and use reacts to respond to their partner's states. The study versions of the app contain a subset of the animations available in the public version, in order to focus on states that could be interpreted from biosignals.

\begin{figure*}[t!]
\centering

  \subfigure[Emotions (sad)]{
      \label{fig:sensed1}
      \includegraphics[width=0.22\textwidth]{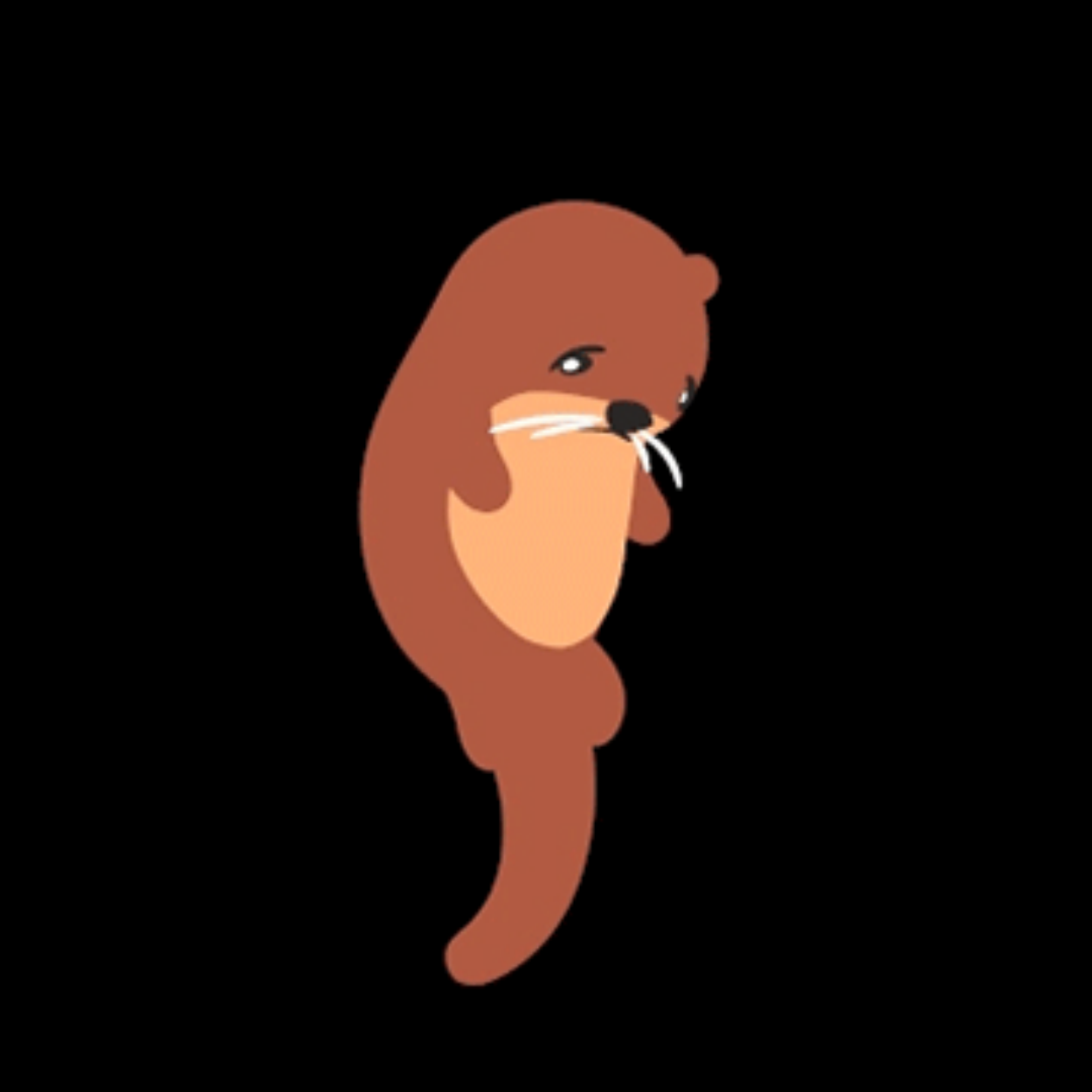}
  }
  \quad
  \subfigure[Activities (sleeping)]{
      \label{fig:sensed2}
      \includegraphics[width=0.22\textwidth]{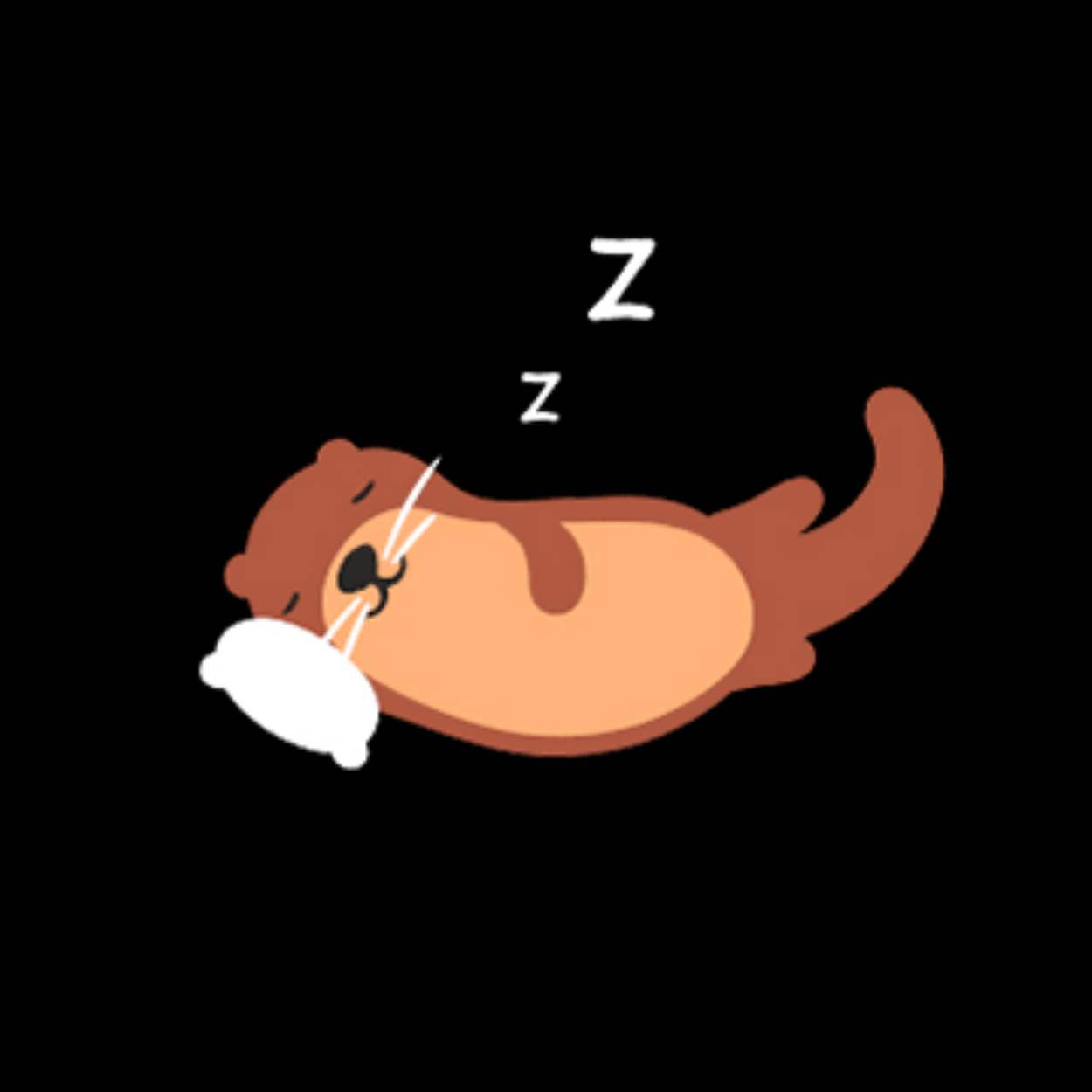}
  }
  \quad
  \subfigure[Greetings (waving)]{
      \label{fig:sensed3}
      \includegraphics[width=0.22\textwidth]{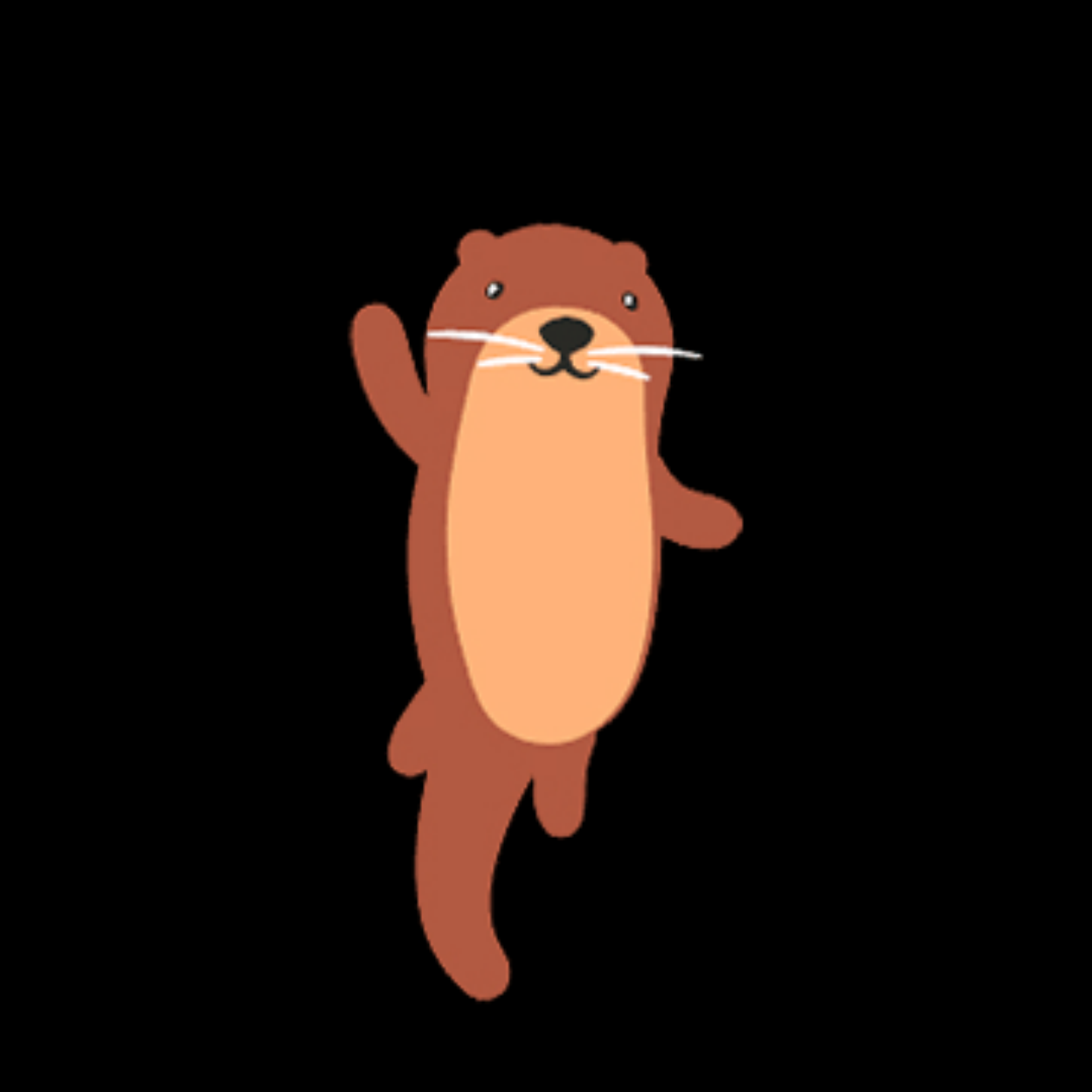}
  }
  \quad
  \subfigure[Affection (handholding)]{
      \label{fig:sensed4}
      \includegraphics[width=0.22\textwidth]{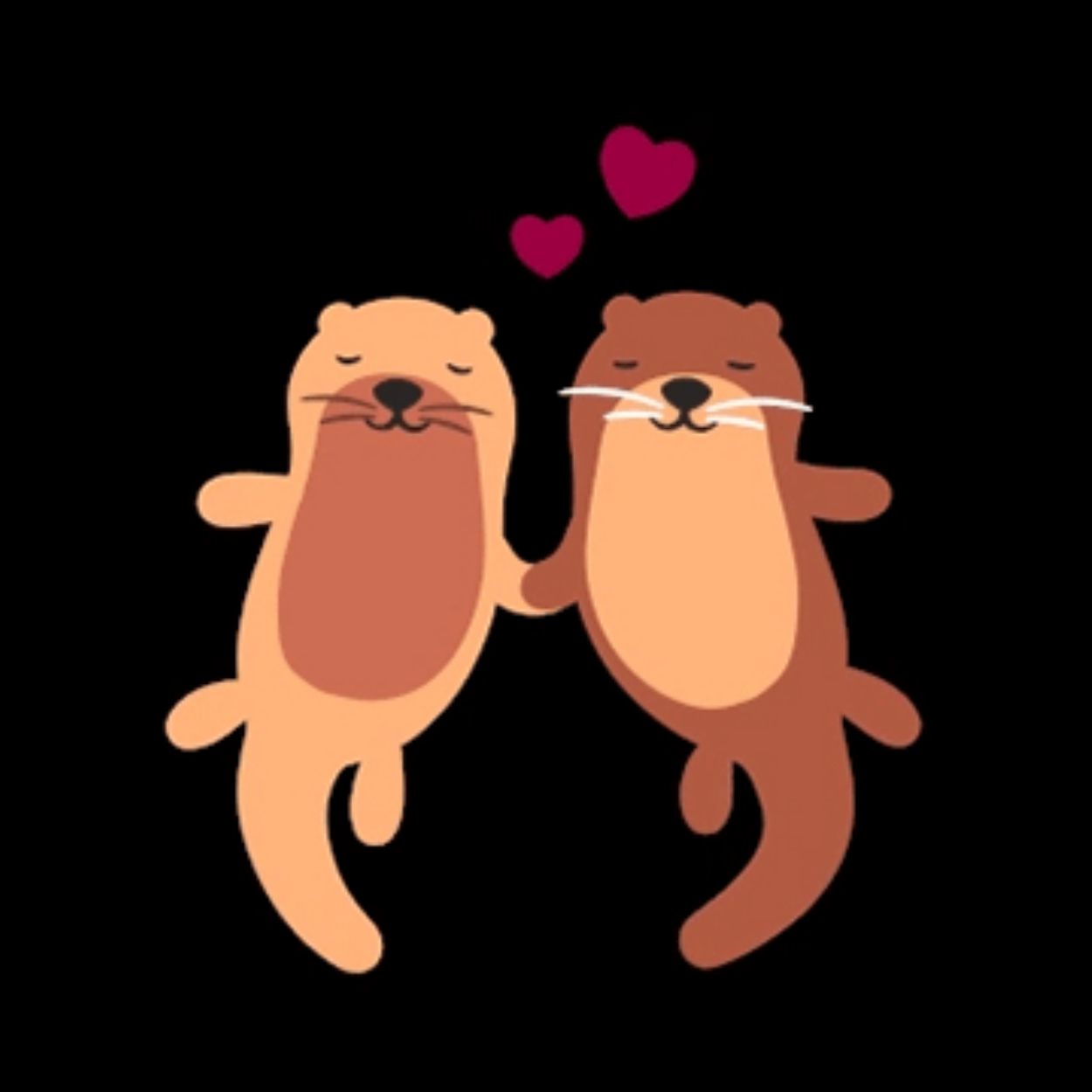}
  }

%   \addtolength{\subfigcapskip}{5pt}
%   \subfigure[Table of states.]{
%     \label{fig:statetable}
  \caption{Examples of state otters.}
  \Description[Screenshots of four otter state animations]{Screenshots of four otter state animations, including a sad, sleeping, and waving otter, and two otters holding hands.}
\end{figure*}

\begin{table}
  \caption{List of state otters.}~\label{fig:states}
    \footnotesize
    \def\arraystretch{1.1}
    \centering
    \begin{tabularx}{\columnwidth}{ X p{2.4cm} p{1.5cm} p{1.5cm} }
      \toprule
      \textbf{Emotions (heart rate)} & \textbf{Activities} & \textbf{Greetings}  & \textbf{Affection} \\
      \midrule
      excited & eating (time) & waving & hugging \\
      calm & sleeping (time) & & handholding \\
      angry & walking (motion) & &  \\
      sad & running (motion) & &  \\
      surprised & exercise (heart rate) & & \\
      bored & & & \\
      neutral & & & \\
      \bottomrule
    \end{tabularx}
\end{table}

\subsubsection{State animations}
Significant Otter presents an \textit{interpreted} representation for biosignals: animated avatars that correspond to different emotional and physical states. That is, the system determines an interpretation for a user's heart rate by mapping it to multiple possible states, as opposed to presenting \textit{raw} biosignals data (e.g., a heart rate number)~\cite{hassib2017heartchat}. We made this decision based on prior work~\cite{liu2017supporting}, which shows that raw data is less engaging and requires additional contextual clarification that may not be feasible on a lightweight smartwatch communication app.

There are four types of state animations: emotions, activities, greetings, and affection. We chose emotion and activity animations according to expressions through biosignals shown in prior work~\cite{liu2017supporting}. We included greeting and affection animations to represent minimal expressions of mutual affection~\cite{hassenzahl2012all}, which prior work suggests are common use cases (``hello,'' ``thinking of you'') on similar apps~\cite{liu2019animo}. To limit the number of available states according to our pilot results, we included only a few greeting and affection states such that we could cover a sufficient number of sensed states (emotions and activities) to address our research questions.

\paragraph{Emotions} These include excited, angry, calm, sad, surprised, bored, and neutral otter animations. We chose these states to represent each quadrant of the valence-arousal model of emotion~\cite{posner2005circumplex}. The sensing ON version senses these states using heart rate data extracted from HealthKit. Since valence cannot be determined from the heart rate data, the system suggests states according to arousal levels determined from the data~\cite{egger2019emotion}. For instance, excited and angry states are available when people are in high arousal, while the calm and sad states are available when they are in low arousal. Significant Otter determines different ranges of heart rate based on people's historical data, including their min, max, walking, and resting heart rates from HealthKit, which Apple updates daily. The ranges are shown in Figure~\ref{fig:hrthresholds}, and were determined through empirical testing within the research team.

\paragraph{Activities} These animations represent daily activities, including eating, sleeping, walking, running, and exercising. Eating and sleeping are time-based activities, inferred from heart rate changes during specific times. Eating is detected based on common meal times in the US (11AM-2PM and 5PM-8PM)~\cite{larson2002dinner} and neutral or high arousal~\cite{sauder2012effect}, while sleeping is detected based on common bedtimes and hours slept in the US (10PM-8AM)~\cite{bedtimes} and low or neutral arousal~\cite{krauchi2001circadian}. Walking and running are motion-based activities, classified by Apple's Core Motion\footnote{Since Apple already provides activity detection for walking and running, we did not use heart rate data to sense those states. However, participants perceived the sensed states as tied to heart rate.}. Exercising is detected solely from heart rate changes, and is presented during high or very high arousal.

\paragraph{Greetings.} This category simply contains waving. This is not sensed, as people may want to greet their partner at any moment. Instead, the app rotates availability with the affection animations such that one can always convey either ``hi'' or ``thinking of you'' to their partner.

\paragraph{Affection.} This category shows animations where the couple's two otters are interacting, including hugging and holding hands. These are not sensed, as people may want to show affection at any moment. Instead, these animations randomly rotate with greetings, as described above.

\begin{figure*}[t!]
\centering
  \subfigure[Acknowledgement (thumbs up)]{
      \label{fig:react2}
      \includegraphics[width=0.25\textwidth]{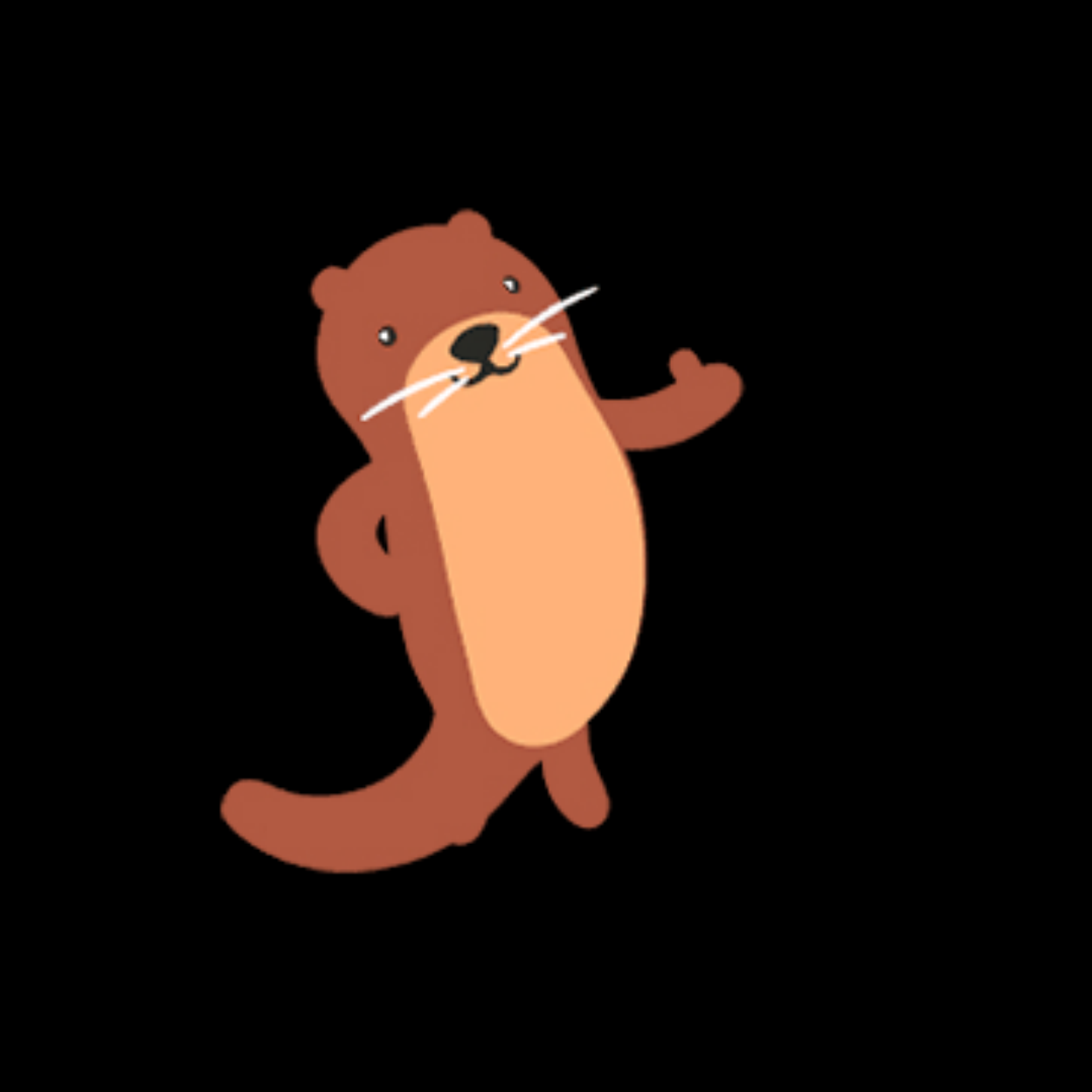}
  }
  \quad
  \subfigure[Caring (pat on the back)]{
      \label{fig:react3}
      \includegraphics[width=0.25\textwidth]{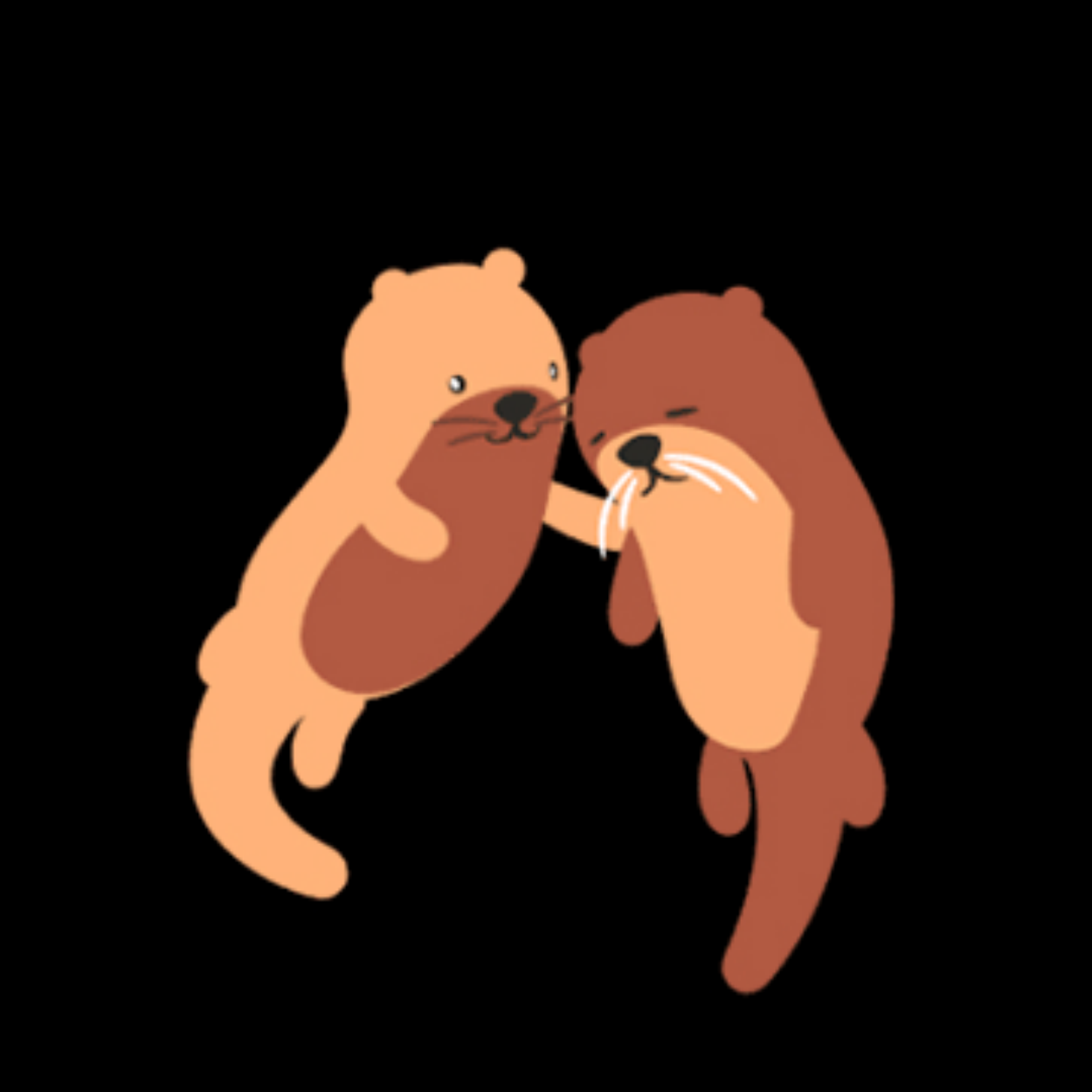}
  }
  \quad
  \subfigure[Follow-up (question)]{
      \label{fig:react4}
      \includegraphics[width=0.25\textwidth]{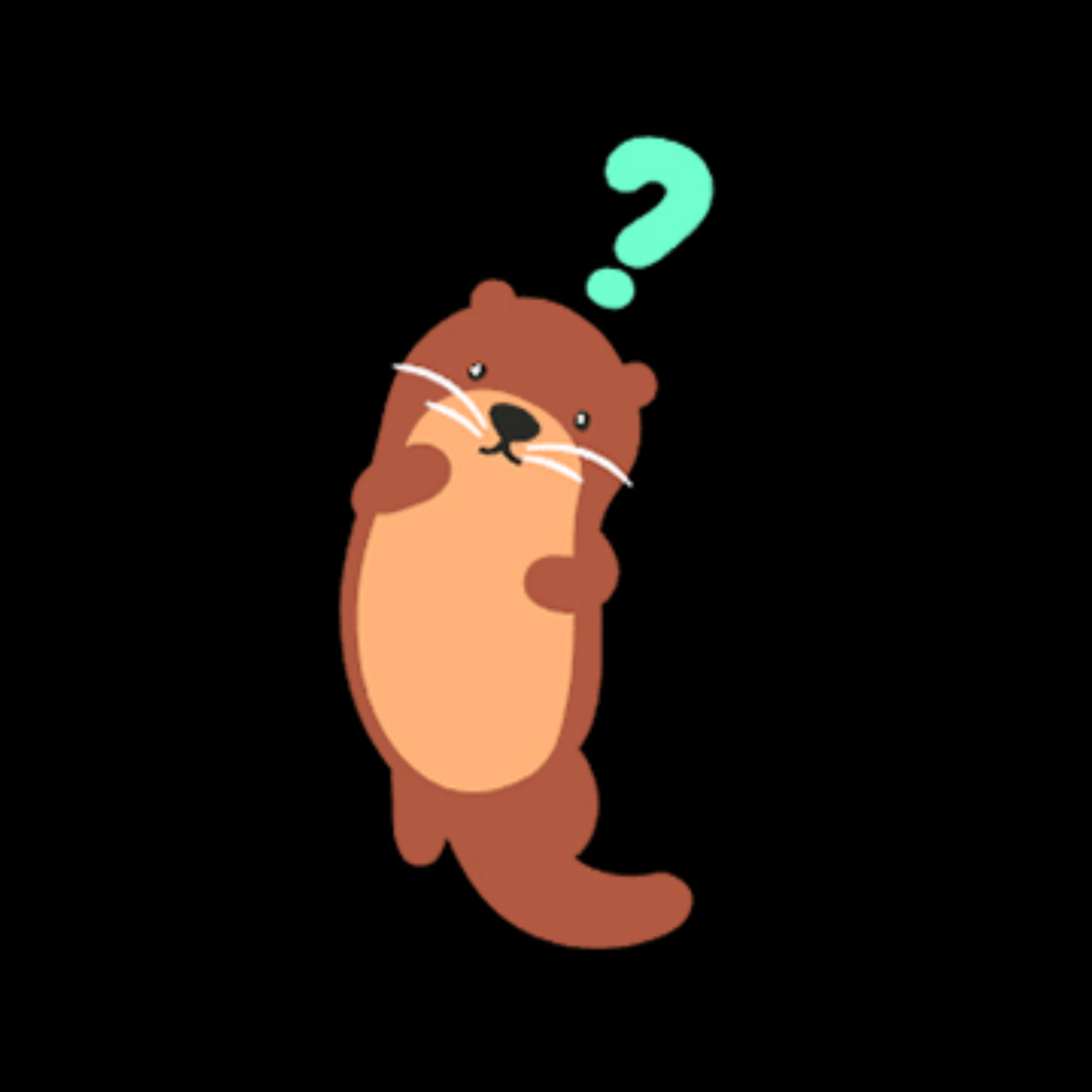}
  }

\caption{Examples of react otters.}
\Description[Screenshots of three otter react animations]{Screenshots of three otter state animations, including a thumbs up and question mark otter, and one otter patting the back of another otter.}
\end{figure*}

\begin{table}
    \caption{List of react otters.}
    \label{fig:reacttable}
    \footnotesize
    \def\arraystretch{1.1}
    \centering
    \begin{tabularx}{\columnwidth}{ p{1.5cm} p{2.5cm} X p{1.5cm} }
      \toprule
      \textbf{Emotions} & \textbf{Acknowledgement} & \textbf{Caring}  & \textbf{Follow-up} \\
      \midrule
      excited & thumbs up & hugging & question \\
      calm & nodding & handholding & call me\\
      angry & & love &  \\
      sad & & pat on the back &  \\
      surprised & & & \\
      bored & & & \\
      \bottomrule
      \end{tabularx}
\end{table}

\subsubsection{React animations} We included 14 different reacts to cover a variety of responses people could have to their partner's state. Significant Otter initially had 22 react animations, which were designed based on social support literature and existing react systems~\cite{tian2017facebook}. For the former, we focused on emotional support, or expressing caring and concern, as other types of support typically require more information~\cite{macgeorge2011supportive,burleson2003emotional,jones201416}, which would not be suitable for a lightweight platform. Reacts are \textit{not} sensed in order to explore people's decisions around responding to their partner's states.

Based on feedback from the first pilot study, we reduced the number of reacts to increase usability. We ran a survey on Mechanical Turk to understand how people perceive the react animations in order to select the most relevant ones. 45 participants interpreted characteristics of each react (e.g., extent of emotional support it provides), gave an example text message that would prompt them to use it as a response, and wrote the response that they believed it conveyed in words (see supplemental materials). We selected the final set of reacts to cover diverse possible responses to the Significant Otter states. This included emotions (similar to states), acknowledging the sender's state or receipt of their state (e.g., ``I agree'' or ``OK''), showing caring and affection for the sender (e.g., ``I'm here for you'' or ``I love you''), and indicating a desire for follow-up on a different platform (e.g., ``Call me ASAP''). We removed reacts that elicited interpretations that did not suit the available states or were overly ambiguous.

\subsection{Sharing Otter States}
People can share their otter state through the main screen of Significant Otter on their watch or phone. People can enter the main screen through notification prompts to view their otter (explained in section~\ref{sec:notif} below), or by opening the app on their own. On the watch, people can open the app through the app complication\footnote{Complications are widgets for the Apple Watch watch face that display information about an app.} on their watch face, which will show an otter icon, similar to an emoji, of one of their currently available states. The icons are designed to act as short-form representations for the larger animations, such that people can glance to see their state otter without opening the app. After opening the app, people can simply tap on their otter to send it to their partner, and their partner can view the otter's animated state on their own device (see Figure~\ref{fig:send_receive}). People can scroll to view and send other possible states, using the Apple Watch crown or swiping up/down on the phone.

We show multiple shareable states due to limitations in detecting emotions from signals available on the watch (e.g., low granularity of heart rate, determining valence), as well as recommendations from prior work~\cite{liu2019animo} to explore systems that collaborate with the user to determine their state. By providing a limited set of other possible states, people can reflect on their subjective emotions alongside the app's suggestions, and then select one of the recommended states. Therefore, the system has low autonomy~\cite{jakesch2019ai}, where the states that people send are partially automated and partially determined by the user. At least one affection or greeting state is always included in the sensed state list, as described above. With sensing OFF, the list is restricted to two to five random states to match the possible sizes of the sensed state list. The states are randomized every 10 minutes to match the frequency of sensing states with sensing ON.

\begin{figure*}[t!]
\centering

  \makebox[1\width][c]{

    \subfigure[Bob is presented with a set of react otters to react to Alice's state otter.]{
      \label{fig:react_flow1}
      {\makebox[1\width][c]{\includegraphics[width=0.2\textwidth]{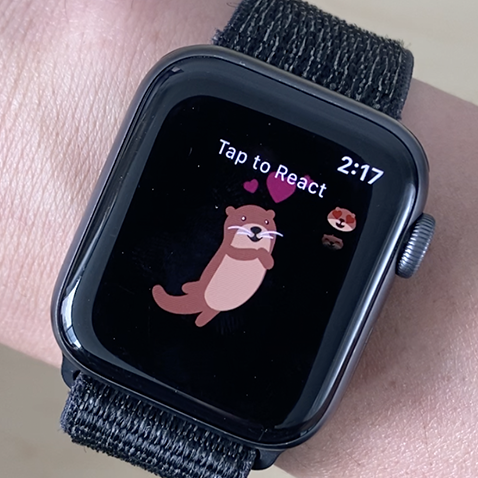}}}
      \hspace{0.2cm}
    }
    \quad
    \subfigure[Bob scrolls through the list of react otters to select one.]{
      \label{fig:react_flow2}
      {\makebox[1\width][c]{\includegraphics[width=0.2\textwidth]{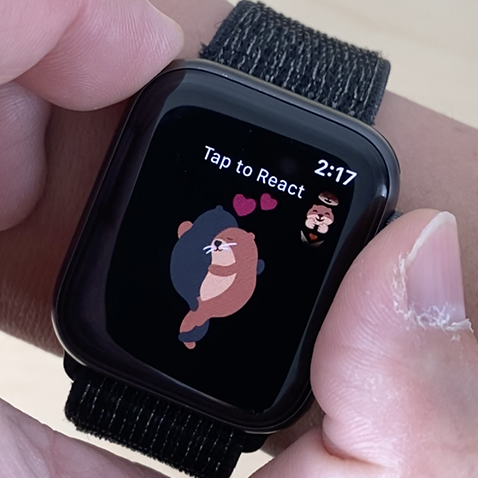}}}
      \hspace{0.2cm}
    }
    \quad
    \subfigure[Bob taps to send his selected react otter to Alice.]{
      \label{fig:react_flow3}
      {\makebox[1\width][c]{\includegraphics[width=0.2\textwidth]{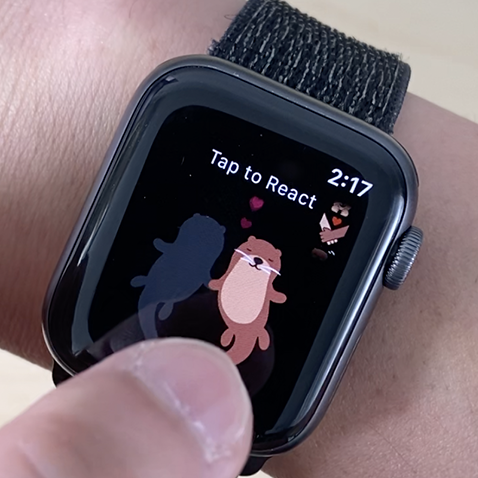}}}
      \hspace{0.2cm}
    }
    \quad
    \subfigure[The app confirms that Bob's react otter was sent to Alice.]{
      \label{fig:react_flow4}
      {\makebox[1\width][c]{\includegraphics[width=0.2\textwidth]{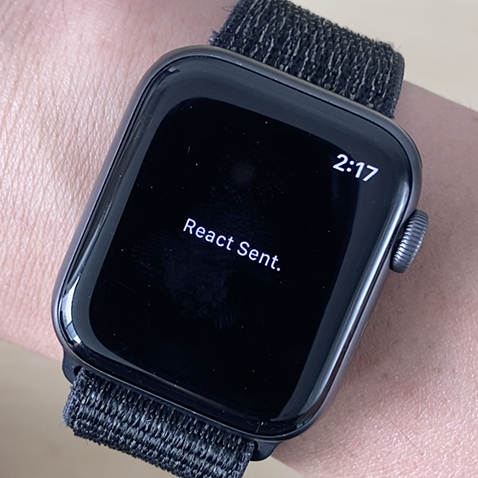}}}
      \hfill
    }
  }
  \caption{Bob reacting to Alice's otter on the Apple Watch.}

  ~\label{fig:react_flow}
 \Description[Screenshots of the user flow for reacting to an otter state]{Screenshots of the user flow for reacting to an otter state on the Apple Watch. A user Bob scrolls through his react otter list. Tapping on an otter sends it to Alice. The app shows text that Bob's React was sent.}
\end{figure*}

\begin{figure*}[t!]
\centering
\makebox[1\width][c]{

  \subfigure[Partner's state otter visit notification with quick reacts (state pictured: sad)]{
    \label{fig:notif_visit}
    {\makebox[1\width][c]{\includegraphics[width=0.2\textwidth]{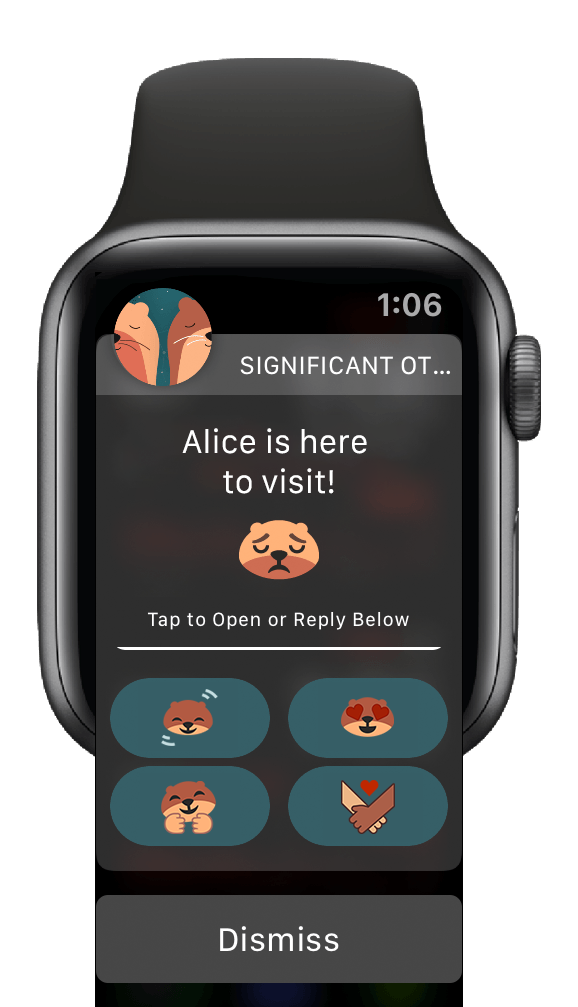}}}
    \hspace{0.2cm}
  }
  \quad
  \subfigure[User's own state otter notification, which comes periodically during the day.]{
    \label{fig:notif_msgfromotter}
    {\makebox[1\width][c]{\includegraphics[width=0.2\textwidth]{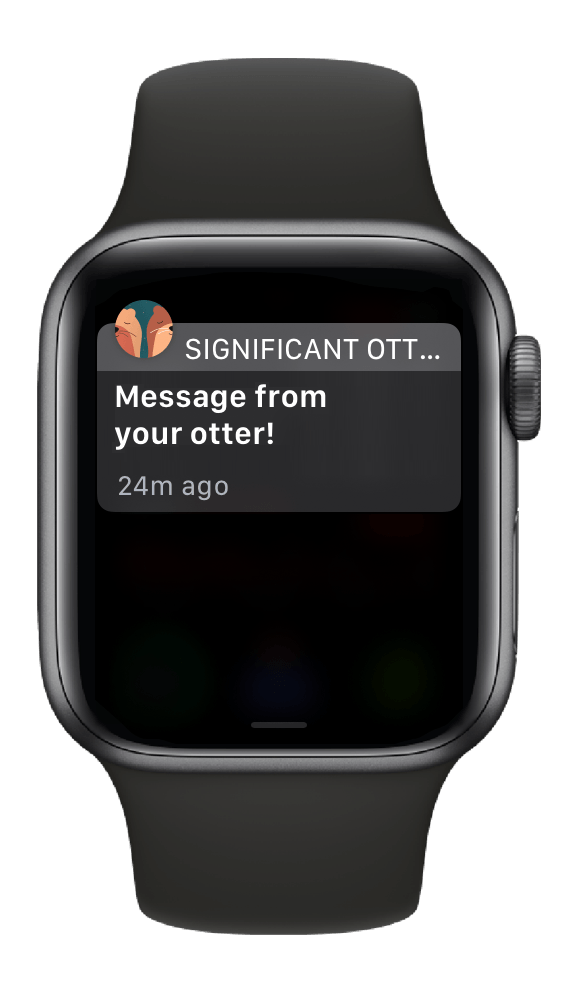}}}
    \hspace{0.2cm}
  }
  \quad
  \subfigure[Opening (b) with sensing ON will show a sensed state icon.]{
    \label{fig:notif_sensed}
    {\makebox[1\width][c]{\includegraphics[width=0.2\textwidth]{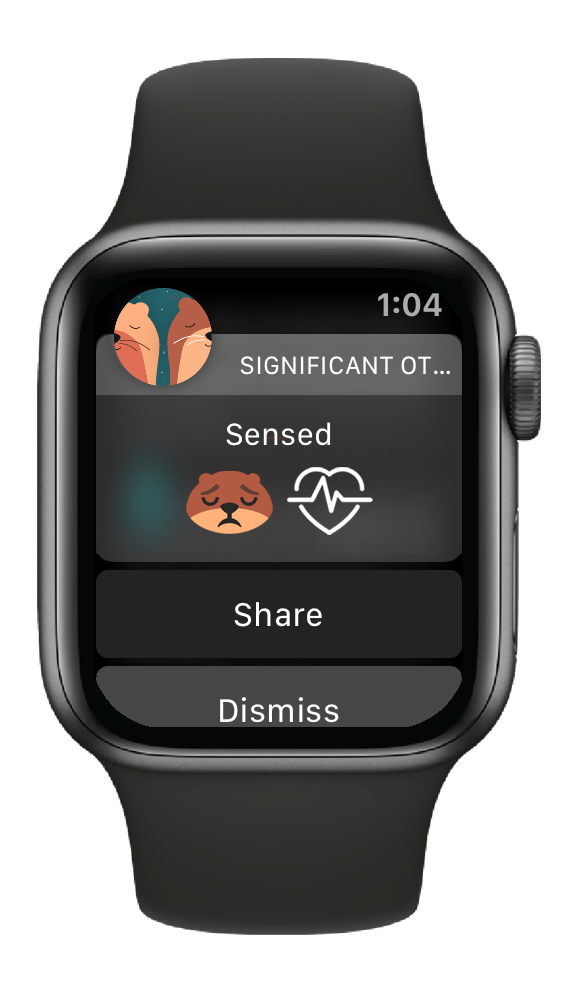}}}
    \hspace{0.2cm}
  }
  \quad
  \subfigure[Opening (b) with sensing OFF shows a randomly available state icon.]{
    \label{fig:notif_random}
    {\makebox[1\width][c]{\includegraphics[width=0.2\textwidth]{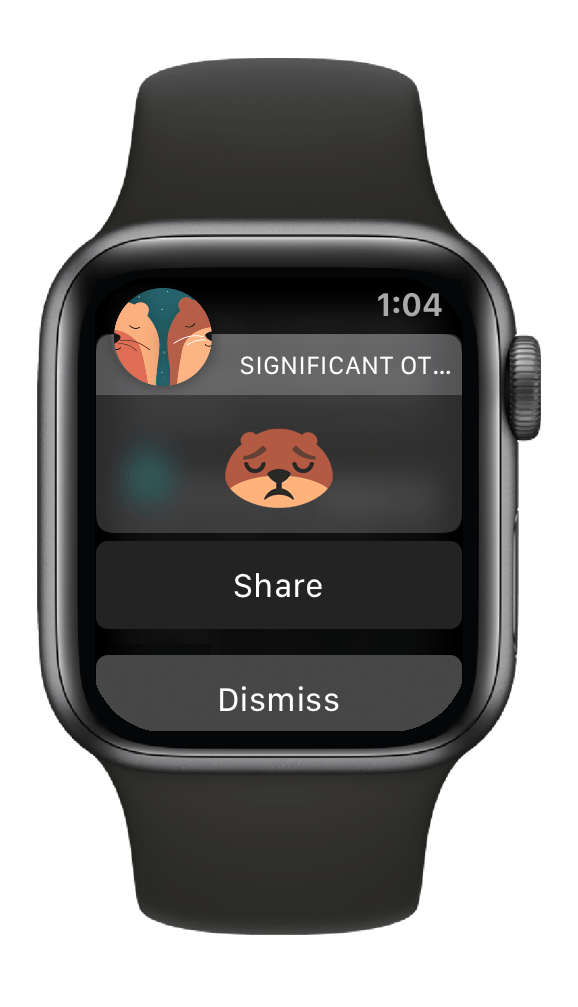}}}
    \hfill
  }
}
\caption{Significant Otter notifications.}~\label{fig:notifications}
 \Description[Screenshots of app notifications]{Screenshots of the app notifications. The partner state otter visit shows a sad otter icon, with four quick react buttons below including nodding, heart-eyed, hugging, and hand-holding otter icons. The user's own state otter notification says ``Message from your otter.'' Opening the state otter notification with sensing ON shows the text ``Sensed'' with a sad otter icon with a heart rate icon with a Share and Dismiss button below. Opening it with sensing OFF just shows the sad otter icon with a Share and Dismiss button below.}
\end{figure*}

\subsection{Reacting to an Otter State}
When people receive their partner's otter, they have the option to react within the app. After they view their partner's state animation, the app will enter the react mode, where they can scroll through all 14 possible reacts (using the watch crown or swiping up/down on the phone) and tap on one to send it as a response to their partner. We include all possible reacts in order to understand people's preferences and decisions in reacting with certain animations. People can also react through ``quick reacts,'' selecting from one of four possible reacts shown in the notification they receive to respond without opening the app. Quick reacts are only featured on the watch and are primarily included for usability, as a lightweight way to respond without viewing the full animation. The four available quick reacts are fixed and selected based on the most frequently used reacts in the public version of the app (love, nodding, handholding, and hugging). Finally, people can choose not to react by selecting ``Don't react'' in the app after viewing the animation, or dismissing the notification. Reacts are the same between both study versions of the app in order to explore potential differences in how people react to their partner's sensed or randomized state. People cannot react to a react animation. After a user views a react, the app will return to the main screen with the list of states.

\begin{table*}[t]
\centering
\caption{Participant couples.}
\label{tab:hedgehug_participants}
\small
\begin{tabular}{llrrlrcccc}
\toprule
\multicolumn{2}{c}{\textbf{participants}} & \multicolumn{2}{c}{\textbf{genders}} &
\multicolumn{2}{c}{\textbf{age}} &
\textbf{relationship length} &
\textbf{married} &
\textbf{cohabitating} &
\textbf{US state} \\
\midrule
P1 & P2 & F & M & 36 & 36 & 1-3 years & No & No & NY \\
P3 & P4 & F & M & 25 & 38 & 1-3 years & No & Yes & MD \\
P7 & P8 & F & M & 22 & 27 & 1-3 years & No & Yes & CA \\
P9 & P10 & M & F & 24 & 24 & 1-3 years & No & No & CA \\
P11 & P12 & M & NB & 24 & 27 & 1-3 years & No & Yes & IN \\
P13 & P14 & M & F & 32 & 37 & > 6 years & Yes & Yes & CA \\
P15 & P16 & M & F & 22 & 20 & 1-3 years & No & Yes & GA \\
P17 & P18 & M & F & 25 & 25 & 1-3 years & No & No & CA \\
P19 & P20 & M & F & 33 & 30 & > 6 years & Yes & Yes & AZ \\
P21 & P22 & M & F & 49 & 51 & 4-6 years & Yes & Yes & OR \\
P23 & P24 & M & F & 33 & 35 & > 6 years & Yes & Yes & TX \\
P25 & P26 & M & F & 20 & 20 & 4-6 years & No & No & MA \\
P27 & P28 & M & F & 26 & 28 & 1-3 years & No & No & CA \\
P29 & P30 & F & M & 24 & 26 & 1-3 years & No & No & VA \\
P31 & P32 & F & M & 32 & 27 & 1-3 years & No & No & CA \\
P33 & P34 & M & F & 22 & 21 & 1-3 years & No & No & AL \\
P35 & P36 & M & F & 19 & 18 & 11 months & No & No & MO / IA \\
P37 & P38 & F & M & 29 & 31 & > 6 years & Yes & Yes & TX \\
P39 & P40 & F & F & 33 & 37 & 1-3 years & No & Yes & TX \\
P41 & P42 & M & F & 25 & 26 & > 6 years & No & Yes & AZ \\ \bottomrule
\end{tabular}
\end{table*}

\subsection{Notifications}\label{sec:notif}
\subsubsection{Partner's state otter visit} People are notified when their partner sends them their state otter. The notification includes an otter icon representing their partner's state otter as well as the four quick reacts (see Figure~\ref{fig:notif_visit}). We used icons for the notifications such that people can glance at the state they received and dismiss or quick react, if desired. Tapping on the icon opens the app to play their partner's state animation, and then enter the in-app react mode.

\subsubsection{Partner's react otter} People are notified when their partner reacts to a state that they sent. The notification includes an icon representing their partner's react otter. Tapping on the icon opens the app to play the animation of the react otter, alongside an icon representing the state to which their partner is reacting.

\subsubsection{User's own state otter} People are notified periodically during the day (at least 45 minutes apart, to minimize invasiveness) with a ``Message from your otter'' notification (see Figure~\ref{fig:notif_msgfromotter}). Opening the notification shows an icon of an available state animation. We include these notifications to encourage people to share their current state with their partner. Notifications are time-based and thus appear \textit{regardless} of whether the user's state changed. This is due to limitations of the Apple Watch, which does not enable real-time heart rate sensing (outside of a fitness tracking session) necessary for recording in-the-moment state changes. This also helped to control differences in the notifications between the two versions of the app, as the app with sensing OFF cannot record state changes. Tapping on the icon allows the user to view the animation in the app. People can also directly send from the notification using the ``Share'' button. With sensing ON, the available state is randomly selected from the list of sensed states (not including greetings
or affection). Sensing ON also shows a heart icon to indicate that the state is selected from the sensed list. With sensing OFF, the available
state is randomly selected from the list of random states.

\section{Methods}
To test Significant Otter, we deployed the sensing OFF and ON versions of the app in a one-month field study.

\subsection{Participants}
Recruitment took place from late March to early April 2020, during the COVID-19 stay-at-home orders for many US states. We recruited 21 romantic couples; however, one couple was removed at the start of the study due to failing to meet the minimum participation requirement (explained in the Procedure below). This left a total of 20 couples (N=40). Participants were recruited through Reddit posts on the SampleSize and AppleWatch subreddits, recruitment posts for about 30 US cities on Craigslist (including major cities in different areas of the US to account for differences in state responses to COVID-19), and snowball sampling (social media posts).

Participants took a screening survey to ensure that they met the study requirements. This included being in an exclusive romantic relationship, living in the US, being able to participate in onboarding and interview sessions via video call, and owning an Apple Watch Series 3 or above that they used for at least two weeks, to ensure familiarity with the watch. Participants described using their Apple Watch for a variety of reasons, predominantly fitness tracking, but also music, news, weather, short texting, and checking notifications. The screening survey also included several questions about participants' circumstances concerning COVID-19, including their living situation with their partner. Results from the second pilot study suggested that people are less likely to engage with the app when collocated with their partner (e.g., if they were both working from home); therefore, we recruited couples who were living apart or living together with one or both of them spending most of their time elsewhere.

The couples we recruited were diverse in several dimensions, including their backgrounds, careers, demographics, and length of relationship. About half of the couples were not living together, including one couple in a long-distance relationship. The rest of the couples were living together with at least one person working outside as an essential worker. Table \ref{tab:hedgehug_participants} summarizes the demographic information per couple. Unfortunately, our sample was not diverse in sexual orientation, where most couples were heterosexual. 13 participants identified as Hispanic, Latino or Spanish, 11 as White/Caucasian, 9 as Asian, 4 as Black/African-American, 1 as Native Hawaiian or other Pacific Islander, 1 as Asian/Hispanic, and 1 as Biracial. Participants had various jobs, including students, healthcare professionals, personal trainers, technicians,  restaurant workers, and analysts. Six were unemployed or furloughed due to COVID-19.

\subsection{Procedure}
We conducted a one-month study deploying Significant Otter in the wild with couples who were Apple Watch users. All participants used the app with sensing OFF for the first two weeks and with sensing ON for the latter two weeks. We intentionally did not counterbalance the order of the versions, as our research questions focused on the shift from the status quo of communicating without biosignals to communicating with biosignals. Moreover, the removal of ``sensing'' as a feature in a counterbalanced study could disrupt participants' mental model of the app, as opposed to a feature update when switching from sensing OFF to  ON. The study consisted of the following sessions:

\subsubsection{Onboarding Session}
Each couple completed a 30 minute onboarding session with one of the researchers over a video call. During this session, participants installed Significant Otter with sensing OFF on their iPhone and Apple Watch through Apple's TestFlight\footnote{TestFlight is an online service for deploying apps to testers.} and added the app as a complication on their watch. During the installation, participants completed a short questionnaire about their background and relationship with their partner. Then, we instructed them on using the app, and asked them to test it during the session to ensure that it was installed and working correctly.

\subsubsection{Daily Usage of Significant Otter}
Participants could freely use Significant Otter as little or as much as they wanted during the study. In order to capture participants' perceptions of the app and behaviors throughout the study, we asked participants via email to complete brief daily surveys about how they used the app that day (see supplemental materials). The first daily survey included comprehension questions to ensure that participants understood how to use Significant Otter. Given the time and effort involved in filling out a survey every day, we required participants to fill out a minimum of three daily surveys per week.

\subsubsection{Mid-Study Session}
Two weeks after their onboarding session, each participant individually completed a 30-60 minute mid-study session over video call. Before the session, participants completed a mid-study questionnaire similar to the onboarding questionnaire, with the addition of questions about COVID-19-related changes they experienced since the start of the study. During the call, we conducted a semi-structured interview about participants' experiences with the sensing OFF version of the app (see supplemental materials). The interview included questions about participants' overall thoughts and perceptions about the app and its different features (e.g., reactions to the notifications), and how they used the app with their partner (e.g., when and why they sent their own otter, what they thought of their partner's otter and how they responded, if at all). To help participants recall their experiences, we showed GIFs of the top states and reacts that they sent and received. After completing the interview, participants installed and were given instructions on the sensing ON version of the app. We explained that with sensing ON, the app displays state animations based on heart rate using Health data from the Apple Watch\footnote{In the session, we stated: ``Version 2 will try to sense your state using your heart rate and other contextual data. The state otters that you’ll be able to see and send to your partner will be based on this sensing.''}.
After the session, participants were given a \$75 Amazon gift card as mid-point compensation.

\subsubsection{Exit Session}
Two weeks after their mid-study session, each participant individually completed a 30-60 minute exit session over video call. Before the session, participants completed a questionnaire identical to the mid-study questionnaire. During the call, we conducted a semi-structured interview about participants' experiences with the sensing ON version of the app. The interview was similar to the mid-study interview, with the addition of questions about how they understood and perceived sensing in the app (see supplemental materials). At the end of the interview, we asked participants to uninstall  Significant Otter. Since multiple participants expressed interest in continuing to use the app, we provided links to the public version, which participants could freely download. After the session, participants were compensated with another \$75 Amazon gift card for completing the study.

\subsubsection{Adjustments and Issues During the Study}
During the study, we made updates to the app and some study materials to address issues that emerged as participants used the app. Participants installed bug fixes by uninstalling and reinstalling the app through TestFlight. We also adjusted how we explained the sensing ON version and added related comprehension questions to the daily surveys, since some participants expressed confusion around how sensing worked (e.g., whether opening the app triggered sensing). Additionally, several participants had issues receiving the app notifications, where they may not have received any during the first or second half of the study. This is a limitation of the Apple Watch OS, which restricts when and how often apps can send push notifications. Since we were unable to address this issue, participants received notifications at different frequencies.

\subsection{Analysis}
We analyzed transcripts of the mid-study and exit interviews using a grounded theory approach~\cite{strauss1998basics}, focused on how participants' perceptions and behaviors shifted between the two versions of the app. First, we segmented the transcripts into high-level categories according to our interview protocol, which highlighted the different aspects of communication (e.g., content of what they sent, thoughts about their partners' sent state, feedback provided, etc.). This enabled us to analyze similar concepts together. Categories were the same for the mid-study and exit interviews, with the addition of categories for the exit interview specific to sensing (e.g., how sensing worked) and external factors that affected usage (e.g., novelty effects, COVID-19-related changes). Next, we developed open codes for each category based on a subset of transcripts, labeling them according to similarities in participants' responses. Three coders validated the subsequent codebook by independently coding another subset of transcripts, meeting frequently to discuss the codes and ensure high inter-rater reliability. They achieved fuzzy Fleiss' kappas~\cite{kirilenko2016inter} above 0.7. After validating the codebook, the three coders divided and coded the rest of the segments. We then performed axial coding by grouping similar codes together and analyzing them to form cross-cutting themes. Finally, we refined these themes according to our research questions.

\section{Results}
In this section, we describe how participants used Significant Otter during the study and provide detailed insights around the reasons behind their usage for both versions of the app. Overall, participants sent a total of 2,474 states and 987 reacts during the study. For both app versions, we observed novelty effects in the first week after installation (weeks 1 and 3)\footnote{In week 3, participants described using the app more due to curiosity around the new sensing feature.}. Participants used the app daily even after novelty effects wore off, sending an average of 1.66 states and 0.71 reacts per day with sensing OFF (week 2) and an average of 1.54 states and 0.54 reacts with sensing ON (week 4). There was a slight non-significant drop in usage from sensing OFF to sensing ON, which participants attributed to COVID-19-related changes to their life circumstances (e.g., changes in their work schedule or ability to meet their partner). Despite using the sensing ON version less, 30 out of 40 participants preferred it over sensing OFF for enhancing their ability to communicate and connect with their partner. At the same time, participants experienced challenges in using the app to communicate what they wanted to their partner. We describe these opportunities and challenges for integrating biosignals into communication in more detail below.

\subsection{Opportunities for Biosignals: ``Enhanced Emojis''}
Our results suggest that sharing sensed states can promote efficient and personal communication between couples, and help them feel connected with each other. This aligns with prior work~\cite{liu2019animo}, which similarly showed that people can easily keep in touch through sharing biosignals-driven animated shapes on their smartwatch. We build on this work through our comparison of participants' usage of Significant Otter with sensing OFF and ON. Specifically, we suggest that the app with sensing OFF functioned similar to emojis, stickers, and GIFs, while sensing ON introduced a new, enhanced form of communication.

\subsubsection{Effects on the Stages of Communication}
\paragraph{Easier sharing} Participants felt that sharing from sensed state suggestions was easier than sharing from a randomized list of animations. With sensing OFF, participants scrolled through the list of state animations as if they were scrolling through a shorter emoji/sticker/GIF keyboard. They appreciated the readily available, unintrusive messages that helped them quickly communicate back and forth with their partner without having to use words. However, some participants were frustrated with access to only two to five random states, expecting a wider and more expressive variety, while others appreciated that they did not need to look through hundreds of options for a specific one. After updating to sensing ON, participants perceived that the state list was personalized to them, where the sensed states were more accurate to how they were feeling or what they were doing than the randomized states. Thus, their otter became more representative of them and was easier to send to their partner.

\begin{quote}
\textit{
``The sensed state otters [with sensing ON] hit closer to home than [with sensing OFF], where they were just random otters. So it narrowed it down better.''} - P3
\end{quote}

This was reinforced by the smartwatch, which participants were more compelled to use in the latter half of the study in order for the sensing feature to work. P14 noted that the sensing feature gave him a reason to use his watch:

\begin{quote}
\textit{
``I'm not really super attached to this watch compared to my phone.... So I think having a reason to want to use the watch for this app and its sensing my vitals and things like that is...a good thing to include in this type of app.''} - P14
\end{quote}

The smartwatch prompted participants with notifications that became personalized suggestions about how they were feeling with sensing ON, rather than dismissable nudges to use the app with sensing OFF. When participants viewed the notifications without opening the app, participants simply made ``yes or no'' decisions to share their otter rather than scrolling through emoji-like options or even thinking to send their otter on their own.

\begin{quote}
\textit{
``Because it knows exactly how you're feeling versus like me having to look through it and kinda of tick something. Because sometimes I don't even know how I'm feeling...you don't really think about how you're feeling until you have to...sit and think about it.''} - P9
\end{quote}

\paragraph{Less ambiguity} Sharing a sensed otter was also less ambiguous than sharing a randomized otter. With sensing OFF, participants assigned various meanings to the otter animations, including those that they were not originally designed for, such as suggestions and needs. Participants would even send animations with no intended meaning, just to send one to their partner. This flexibility in the otter animations aligns with the flexibility of emojis, where an emoji can be used to convey numerous possible messages~\cite{wiseman2018repurposing,miller2016blissfully}, and are thus known to be expressive yet ambiguous even when used in textual contexts~\cite{miller2017understanding}. With sensing OFF, some participants described following-up over verbal conversation to clarify the animations they sent, or struggling to interpret and respond to the animations they received, subsequently resorting to ``safe'' otter reacts.

\begin{quote}
\textit{
``Like that walking otter...she might [react with a] thumbs up, but that doesn't necessarily mean that...she's understanding that the walking otter means that I would like to go for a walk...the thumbs up could be, `yes, I want to go for a walk too.' It could be like, `walking is good.' So I mean, there's not a lot of clarity with just having the simplistic reaction.''} - P38
\end{quote}

\begin{quote}
\textit{
``I was confused about that otter...out of my confusion I replied with him because he looked pretty chill.... I was just trying to say no hostility as well, 'cause I didn't know what the other otter was doing. So that was a pretty safe response.''} - P33
\end{quote}

Our results suggest that biosignals helped reduce this ambiguity, where participants no longer assigned different meanings to the animations. Instead, the animations became meaningful on their own, where participants understood them as simply representative of their own or their partner's current state. This facilitated sending state animations, because participants no longer had to think about what they could mean and, like in prior work~\cite{liu2017supporting}, sent them primarily to share their current state. This also facilitated understanding and responding to those animations, because they could appropriately react when they understood what they meant, such as by agreeing, reciprocating, or showing concern for their partner's state.

\begin{quote}
\textit{
``When I was sending the message, my intent behind it wasn't, `Oh, I'm reminding you to do something.' I'm sharing my mood with you...[sensing OFF] was just more inclined to him and [sensing ON] was more inclined to me.''} - P36
\end{quote}

\begin{quote}
\textit{
``I think it's different insofar that I felt like she really wasn't sending random ones. I feel like they were more based on what she was doing. So for that reason I felt like my reactions were more consistent.''} - P39
\end{quote}

\subsubsection{Effects on Social Connection}
\paragraph{Authenticity} Participants felt that sharing sensed otter animations enabled more open and genuine communication with their partner. While couples used both versions of the app to keep in touch with each other's current state, they felt that sensing ON enabled a more personal experience with each other because it was backed by data. Participants described feeling more connected to both their own and their partner's otter because they were tied to their bodies' physical states, as if they were the otters themselves.

\begin{quote}
\textit{
``It's personal to me because it's reading what I'm doing...it's almost as if you could go through the phone yourself and wave or something like that....[With sensing OFF, it] could have just been a sticker app, an iMessage where you're just sending from a collection of animated stickers. Once it [sensed] what you were doing throughout the day, it [became] a more personal experience...because it's sensing what you're wanting to say throughout the day.''} - P35
\end{quote}

\begin{quote}
\textit{
``It was interesting that both of our body's responses were being recorded. That's what I mean by feeling connected like we're not physically together, but you're still able to get a sense of their actual bodily responses through the app, like through technology, and that was cool.''} - P31
\end{quote}

Seeing their own sensed state also encouraged them to reflect on how they were feeling, and be more honest with both themselves and their partner by sharing it with them. Participants similarly felt that their partner was conveying their honest state with them. Some couples noted that even if they are fairly open with each other, they appreciated knowing that their partner's state was backed by data and that their partner was not just putting up a front.

\begin{quote}
\textit{
``I'm pretty open with my feelings overall in life and with my partner, [but with Significant Otter] I'm more open to be like, honest, I guess, like totally 100\% honest compared to 95\% honest...the 5\% can sometimes make a big difference.... I would send the [stressed otter] instead of being like, `Oh, I don't want to look weak right now by showing that I am stressed.'''} - P14
\end{quote}

\begin{quote}
\textit{
``I feel like [with sensing OFF]...I wouldn’t know if that was actually how he was feeling or if he just picked [a smiley one]...just to send something nice. So knowing that he actually felt that way and [was] probably a little bit happy...was good.''} - P37
\end{quote}

This motivated a few participants to be more thoughtful and responsive in their reactions to their partner, such as reacting more quickly or deliberately. For example, P31, who tended to not respond immediately to her partner's state with sensing OFF, felt more urgency to respond with sensing ON:

\begin{quote}
\textit{
`` I feel like [the otter's] a way of him reaching out. So for me to just wait [to] respond and not really think much of it, it feels rude not to validate whatever he sent out, because that is...like a virtual extension of him. So I felt like I needed to respond to it as soon as I saw.''} - P31
\end{quote}

P36, who frequently used the quick reacts with sensing OFF, used them less with sensing ON. She explained that since her partner was sharing his emotions with her, she should put more effort into reacting:

\begin{quote}
\textit{
``Although [quick reacts were] very convenient, I just felt more of a responsibility this time to [open the app]. Just because I felt like my partner was sending me state otters off his emotion. [Doing] a quick react otter...it was kind of dismissing the notification in a sense. Opening up the app and scrolling through all reacts so I could choose the right one made me feel like I was more connected with my partner in the interaction.''} - P36
\end{quote}

\subsection{Challenges for Biosignals: Me vs the System}
As a system that recommends a user's current state, Significant Otter with sensing ON experienced challenges in how participants perceived and trusted the sensed states. Though most participants felt that their sensed otter accurately reflected their feelings and activities, six participants were skeptical of the system's ability to sense states and disagreed with the suggestions they saw. The sensed states were also restrictive, where the participants believed they were less likely to find an animation they wanted to send, while randomization presented equal probability of seeing all states. These participants stated that they would have preferred a list of states with more variety to choose from.

\begin{quote}
\textit{
``Whenever it would send something I would usually get the same otters. So I wish when it was sensing something I would be able to get a variety of otters at different times...compared to just one all the time.''} - P29
\end{quote}

\subsubsection{Effects on the Stages of Communication}
\paragraph{Subjective understandings of sensing and emotions} While perceptions of inaccuracy were a barrier for some participants, participants ranged in their definition of accuracy, which affected their ability to send their own otter and understand their partner's otter. Some participants were accepting of the system's small set of possible sensed states, and gave the app room for error. They did not expect the sensed states to be 100\% accurate and reasoned why the app would suggest states that did not quite match them, based on their knowledge of their heart rate or physical state. These participants also described typically being satisfied with at least one state in the list of suggested states, and did not mind if the other states did not fully match them.

\begin{quote}
\textit{``I would say most of the time it definitely matched. I mean it was pretty much on par with what I was doing, which was really cool to see. Sometimes after I would go for a walk or something, it would tell me I was surprised rather than like working out. But it was like a walk, so I was not doing crazy workouts. So I could see how that happened.''} - P34
\end{quote}

\begin{quote}
\textit{
``Most people at any given time throughout the day, you might be feeling a lot of different things...at least with the sensed version...at least one of the things that it was showing you [matched].''} - P38
\end{quote}

Conversely, participants who perceived the app as inaccurate tended to expect exactly one accurate state, giving less flexibility for the app to suggest other states that may not match their feelings. These discrepancies in perceptions of accuracy appear to stem from participants' different lay understanding of emotions and how they relate to heart rate. For instance, P1 was confident in knowing her own feelings better than the app, while P15 was conflicted on whether to follow his body (what the app suggested to him) or mind (how he thinks he feels). On the other hand, P14, quoted above, felt his heart was an indicator of how he truly felt, as opposed to how he thought he felt in his mind.

\begin{quote}
\textit{
``I feel like \textbf{I} know what I feel like...this thing is guessing how I feel based on I don't know what my heartbeat or...I don't think that's accurate.''} - P1
\end{quote}

\begin{quote}
\textit{
``I just never knew which one was the most accurate.... I really just thought it was that first one [in the state list and] the other ones could have been random [or] maybe a second best choice. I just didn't really figure out which one to go with, you know? I'm like betraying my heart rate if I choose a different one [than the first one] or something.''} - P15
\end{quote}

\subsubsection{Effects on Social Connection}
\paragraph{Agency and effort in communicating feelings} On the other extreme, a few participants described blindly trusting the system and sending their otter from the sensed state notifications even if they did not know which animation they were sending. Though the sensing feature was not intended to be highly accurate, these participants felt the system knew their feelings better than they did, and helped them to convey those feelings to their partner.

\begin{quote}
\textit{
``The first two or three times [that I got the sensed notifications] they were on target as far as...the way I was feeling…. That would be the otter that I was trying to send anyway...so once I realized that that's what was happening, I wouldn't think too much about it anymore as far as opening up the app and seeing if it was the otter I wanted to send or anything like that.''} - P23
\end{quote}

One participant warned against this ``power of suggestion,'' where the system could influence them into thinking they felt a certain way. This aligns with prior work by Hollis and colleagues~\cite{hollis2018being}, which suggests that people may overly trust emotion sensing systems and be influenced by the system's interpretations of their emotions.

\begin{quote}
\textit{
``With [sensing ON] it was like always asking yourself whether or not you really felt that way before sending it. And so I don't know if sometimes that would influence you to send it anyways or influence you to maybe feel that way. Yeah so, I do prefer [sensing OFF,] that way whatever you're feeling...you're able to just think of it on your own and just send it.''} - P18
\end{quote}

Another participant noticed that by simply accepting and sharing the system's recommendation, he put less thought into curating a message to send to his partner. He pointed out that while the message itself was personalized to himself, he no longer took the time and effort to consider what to send:

\begin{quote}
\textit{
``I think with [sensing ON] it was me sending stuff based on what I think the watch read that I felt. So it wasn't me taking the time and going through and saying, yeah, this is the one. It was like, the watch said this is how I feel. So I guess this is how I feel. Let me send it. [It] was almost more impersonal, even though it was reading off of my data.''} - P2
\end{quote}

Though the reduced effort made keeping in touch easier, effort is an important quality of communication that contributes to meaningful connection and close relationships~\cite{kelly2017demanding,kelly2018s}. Moreover, work on AI-mediated communication suggests that systems that generate messages for communication, such as Significant Otter's sensed state animations, can affect perceptions of authenticity~\cite{monroy2011computers} and trustworthiness~\cite{jakesch2019ai} in the sender of the message. Thus, despite sensed states being inherently more personal and intimate, they could potentially prompt less personal ways of communicating if the system has more agency than the user. In the following section, we recommend future research directions and system designs to explore how to reconcile this tension.

\section{Discussion}
\subsection{Summary of Results}
Overall, participants viewed both versions of Significant Otter as a lightweight communication channel that enabled them to easily to keep in touch with their partner and let each other know how they are doing. In their baseline usage of the app with sensing OFF, participants felt the otter animations were an easy way to communicate without words, using them to convey their current state and suggest activities. However, this communication was also limited because a simple animation could mean multiple things or required more detail.

For most participants, Significant Otter with sensing ON mitigated some of the issues with sensing OFF and enhanced participants' ability to connect with each other. Our results show that biosignals can facilitate each stage of communication (\textbf{RQ1}), where sending, understanding, and responding to the state animations were easier with sensing ON than with sensing OFF, because the app curated more straightforward messages for them to send, distilled to participants' sensed states as opposed to user-prescribed meanings. Biosignals also supported feelings of connection between romantic partners (\textbf{RQ2}), where participants described feeling compelled to share honest emotions through the sensed state animations, rather than put up a front. Additionally, they felt the sensed otters were more representative of them and their partner, conveying a sense of their body's physical responses. Taken together, our results describe the \textit{role} of biosignals in dyadic communication, distinct from system features such as the smartwatch or animated avatars~\cite{liu2019animo}, as promoting easier and more authentic communication. However, despite these benefits, biosignals also introduced new concerns around accuracy and agency over the message, where some participants felt the system was overly suggestive on how they were feeling or what to communicate to their partner. This further illustrates the tensions in AI-recommended versus subjectively interpreted emotions found in prior work~\cite{hollis2018being,howell2018tensions}, particularly due to variations in lay understanding or confidence around one's own emotions.

\subsection{Design Implications and Future Directions}
\subsubsection{Sharing sensed states on existing platforms}
While having a separate platform like Significant Otter dedicated to sharing states can create an intimate experience for couples, the sensed otter animations could easily integrate with existing platforms as ``enhanced emojis.'' People already increasingly need to navigate multiple communication apps, which can cause ``expression breakdowns'' when they are unable to consistently express themselves across those apps~\cite{griggio2019customizations}. By integrating biosignals into existing platforms, people could benefit from centralized communication with their partners while expressing themselves in more authentic ways through the sensed states. In platforms such as texting and mobile messaging, they could also could easily start new conversations about the states they share, a common pattern we observed in our study.

As part of existing platforms, biosignals would primarily function as a means to augment communication as opposed to acting as standalone messages like in Significant Otter. Rather than relying on relationship context, people would reference the augmented communication content to interpret the biosignals (e.g., text in mobile messaging~\cite{liu2017supporting,hassib2017engagemeter}). Researchers and designers of communication platforms could explore how biosignals could augment various types of communication content, such as images, videos, or emojis, and new interaction patterns that may emerge. For example, biosignals could become new types of ``emojis'' or integrate with existing emojis by suggesting specific emojis or limiting the available options. This could help people navigate the ever growing list of emojis, as well as clarify potentially ambiguous emojis. Suggested emojis could be annotated in order to designate them as sensed states (e.g., a heart symbol, beats per minute, or special effect or badge attached to the image).

\subsubsection{Addressing user expectations for sensing}
We found that varying perceptions of accuracy and agency over the animations affected participants' ability to use the sensing ON version of the app. Given people's own subjective understanding of their state as well as ongoing research on emotion detection, designers need to consider how to present and incorporate sensing technology both in its current and future levels of accuracy. That is, even if the system claims to be accurate based on the user's physical state, the detected emotion may conflict with how the user subjectively believes they feel~\cite{liu2019animo} or overly influence their feelings~\cite{howell2018tensions,hollis2018being}. For Significant Otter, we lowered the autonomy of the system~\cite{jakesch2019ai}, where the app suggested both a single state in notifications and a list of possible states within the app. Some participants continued to be skeptical of the suggested states, having strong beliefs about how they are feeling, while others were confused by our design, believing that they should see only one recommendation. In fact, a few people did focus solely on the one recommendation in the notification, ignoring the list of in-app states.

Future directions in this area should investigate new ways for expressive biosignals systems to collaborate with people's subjective understanding of their own state. For example, researchers could explore systems that support different lay theories of emotions and how they affect perceptions of the system's accuracy, such as whether the user interprets emotions based on external contextual cues or internal physiological experiences~\cite{uchida2009emotions,zammuner2000men}. Designers of these systems should clearly and carefully introduce how sensing in the system works. For example, onboarding steps could detail the system's approach to emotion (e.g., its relationship to the body's physiological state, why the system might suggest multiple possible emotions), or provide adjustable settings that match one's personal understanding of their own state. Future work could also explore how to involve people in system recommendations, such that they can have more control over what they are feeling and how they share those feelings. For example, the system could allow people to provide feedback on their state in order to improve the system and feel involved in the system's suggestions, or prompt them to interpret the suggested state before sharing it with their partner. This could encourage people to engage in more effort and meaning-making with their partner, and enhance the authenticity of the AI-recommended message.

Finally, researchers should consider how people's understanding of their state and expectations for the system may be affected by different types of expressive biosignals. People may have a more developed understanding about their heart rate given its accessibility, including in fitness trackers and watches, or simply being able to feel when one's own heart beats faster. On the other hand, less accessible biosignals such as skin conductance or brain activity may be less understood or produce different lay interpretations (e.g., associating the brain with cognitive functions), which could potentially affect people's willingness to accept the system's state recommendations.

\section{Limitations}
Though our findings elucidate the value of expressive biosignals in communication, there are several limitations to this work. First, we ran a non-counterbalanced within-subjects study in order to reduce confusion in participants' mental model of the app, where sensing was a ``feature update'' rather than a feature being removed. Most participants perceived sensing ON as a feature update that enhanced their use of the app; however, a few were strongly influenced by their mental model of the app with sensing OFF and expected it to work the same way. Moreover, novelty effects were much stronger for sensing OFF than for sensing ON. The number of sent messages dropped by 605 messages between the first and second week of using sensing OFF, compared to a drop of 77 messages between the first and second week of using sensing ON. Many participants also described getting used to the app during the second half of the study. We took these differences into consideration during both our interviews and analyses; however, future work should consider a between-subjects design or longer longitudinal study to reduce potential order or novelty effects.

Second, we deployed the app \textit{in situ} on participants' own smartwatches for use in their everyday lives, in order to achieve high ecological validity. Given the differences in participants' lifestyles, especially during the COVID-19 pandemic, as well as tendencies towards different devices (e.g., participants with large hands experiencing difficulty interacting with the watch app interface), participants naturally had diverse experiences with the app. Additionally, the app had hard-coded times set for certain states, which may not have matched people's typical patterns. Future systems should consider providing options for people to set their typical meal or sleeping times, or sense additional signals such as location to better predict their state. Limitations of the Apple Watch OS also affected whether the app worked as intended for all participants, where some participants received no notifications while others felt that they received too many. Thus, while our qualitative findings present a variety of interesting communication patterns that stem from participants' diverse usage, studies with greater levels of control or larger sample sizes are necessary to clarify potential causal effects that biosignals have on communication. More granular data collection would also help to capture and further understand the differences in participants' usage, such as the number of notifications that influenced sending a state or participants' perceptions of each sent state.

Finally, while we recruited a diverse sample of participants from different backgrounds, participants were self-selected and may have shown a greater interest in wearable and couple-specific technologies. Additionally, the shortest relationship length among participants was 11 months. People in earlier stages of their relationship or without established communication practices with each other may use the app in different ways. Given stay-at-home orders, we also restricted recruitment to people who were living apart from their partner or living together if one or both of them were essential workers. Thus, we were unable to capture how people that did not match these criteria might use the app outside of these unusual circumstances. It is also possible that our participants would engage with the app differently outside of these circumstances, as many had to adjust to changes in their daily routine during the study.

\section{Conclusion}
We ran a month-long within-subjects field study on Significant Otter, an Apple Watch and iPhone app that enables biosignals sharing through animated otter messages, to explore the role of biosignals in communication. We compared participants' usage of Significant Otter with and without biosignals. Results showed that biosignals can support easier and more authentic communication, while eliciting concerns around accuracy and agency over the communication content based on participants' diverse understandings of emotions. We provide insights on the opportunities and challenges around integrating biosignals into communication
and make recommendations for future research and design, including applying biosignals to existing communication platforms to promote open communication as ``enhanced emojis,'' and exploring greater user involvement in AI-recommended states.

\section*{Acknowledgements}
We would like to thank John Tang and Mayank Goel for their early suggestions that helped shape our study design, as well as their feedback on our results. We are also grateful to David Lin for his help with our qualitative analysis, and Sven Kratz for his feedback on our paper submission.

\bibliographystyle{ACM-Reference-Format}
\bibliography{references}


\begin{thebibliography}{69}

%%% the Web address.

\ifx \showCODEN    \undefined \def \showCODEN     #1{\unskip}     \fi
\ifx \showDOI      \undefined \def \showDOI       #1{#1}\fi
\ifx \showISBNx    \undefined \def \showISBNx     #1{\unskip}     \fi
\ifx \showISBNxiii \undefined \def \showISBNxiii  #1{\unskip}     \fi
\ifx \showISSN     \undefined \def \showISSN      #1{\unskip}     \fi
\ifx \showLCCN     \undefined \def \showLCCN      #1{\unskip}     \fi
\ifx \shownote     \undefined \def \shownote      #1{#1}          \fi
\ifx \showarticletitle \undefined \def \showarticletitle #1{#1}   \fi
\ifx \showURL      \undefined \def \showURL       {\relax}        \fi
% invisible to TeX
\providecommand\bibfield[2]{#2}
\providecommand\bibinfo[2]{#2}
\providecommand\natexlab[1]{#1}
\providecommand\showeprint[2][]{arXiv:#2}

\bibitem[\protect\citeauthoryear{Aron, Aron, and Smollan}{Aron
  et~al\mbox{.}}{1992}]%
        {aron1992inclusion}
\bibfield{author}{\bibinfo{person}{Arthur Aron}, \bibinfo{person}{Elaine~N
  Aron}, {and} \bibinfo{person}{Danny Smollan}.}
  \bibinfo{year}{1992}\natexlab{}.
\newblock \showarticletitle{Inclusion of Other in the Self Scale and the
  structure of interpersonal closeness.}
\newblock \bibinfo{journal}{\emph{Journal of Personality and Social
  Psychology}} \bibinfo{volume}{63}, \bibinfo{number}{4}
  (\bibinfo{year}{1992}), \bibinfo{pages}{596}.
\newblock
\urldef\tempurl%
\url{https://doi.org/10.1037/0022-3514.63.4.596}
\showDOI{\tempurl}


\bibitem[\protect\citeauthoryear{Bales, Li, and Griwsold}{Bales
  et~al\mbox{.}}{2011}]%
        {bales2011couplevibe}
\bibfield{author}{\bibinfo{person}{Elizabeth Bales}, \bibinfo{person}{Kevin~A
  Li}, {and} \bibinfo{person}{William Griwsold}.}
  \bibinfo{year}{2011}\natexlab{}.
\newblock \showarticletitle{CoupleVIBE: mobile implicit communication to
  improve awareness for (long-distance) couples}. In
  \bibinfo{booktitle}{\emph{Proceedings of the ACM 2011 Conference on Computer
  Supported Cooperative Work}}. ACM, \bibinfo{pages}{65--74}.
\newblock
\urldef\tempurl%
\url{https://doi.org/10.1145/1958824.1958835}
\showDOI{\tempurl}


\bibitem[\protect\citeauthoryear{Bentley and Metcalf}{Bentley and
  Metcalf}{2007}]%
        {bentley2007sharing}
\bibfield{author}{\bibinfo{person}{Frank~R Bentley} {and}
  \bibinfo{person}{Crysta~J Metcalf}.} \bibinfo{year}{2007}\natexlab{}.
\newblock \showarticletitle{Sharing motion information with close family and
  friends}. In \bibinfo{booktitle}{\emph{Proceedings of the SIGCHI Conference
  on Human Factors in Computing Systems}}. \bibinfo{pages}{1361--1370}.
\newblock
\urldef\tempurl%
\url{https://doi.org/10.1145/1240624.1240831}
\showDOI{\tempurl}


\bibitem[\protect\citeauthoryear{Boomerang}{Boomerang}{2020}]%
        {boomerang}
\bibfield{author}{\bibinfo{person}{Boomerang}.}
  \bibinfo{year}{2020}\natexlab{}.
\newblock \bibinfo{title}{Respondable: Write Better Email}.
\newblock
\newblock
\urldef\tempurl%
\url{https://www.boomeranggmail. com/respondable/}
\showURL{%
\tempurl}


\bibitem[\protect\citeauthoryear{Burleson}{Burleson}{2003}]%
        {burleson2003emotional}
\bibfield{author}{\bibinfo{person}{Brant~R Burleson}.}
  \bibinfo{year}{2003}\natexlab{}.
\newblock \showarticletitle{Emotional support skills}.
\newblock \bibinfo{journal}{\emph{Handbook of Communication and Social
  Interaction Skills}} (\bibinfo{year}{2003}), \bibinfo{pages}{551--594}.
\newblock
\urldef\tempurl%
\url{https://doi.org/10.4324/9781410607133}
\showDOI{\tempurl}


\bibitem[\protect\citeauthoryear{Cho, Oh, Kim, and Lee}{Cho
  et~al\mbox{.}}{2020}]%
        {cho2020share}
\bibfield{author}{\bibinfo{person}{Hyunsung Cho}, \bibinfo{person}{Jinyoung
  Oh}, \bibinfo{person}{Juho Kim}, {and} \bibinfo{person}{Sung-Ju Lee}.}
  \bibinfo{year}{2020}\natexlab{}.
\newblock \showarticletitle{I Share, You Care: Private Status Sharing and
  Sender-Controlled Notifications in Mobile Instant Messaging}.
\newblock \bibinfo{journal}{\emph{Proceedings of the ACM on Human-Computer
  Interaction}} \bibinfo{volume}{4}, \bibinfo{number}{CSCW1}
  (\bibinfo{year}{2020}), \bibinfo{pages}{1--25}.
\newblock
\urldef\tempurl%
\url{https://doi.org/10.1145/3392839}
\showDOI{\tempurl}


\bibitem[\protect\citeauthoryear{Counts and Fellheimer}{Counts and
  Fellheimer}{2004}]%
        {counts2004supporting}
\bibfield{author}{\bibinfo{person}{Scott Counts} {and} \bibinfo{person}{Eric
  Fellheimer}.} \bibinfo{year}{2004}\natexlab{}.
\newblock \showarticletitle{Supporting social presence through lightweight
  photo sharing on and off the desktop}. In
  \bibinfo{booktitle}{\emph{Proceedings of the SIGCHI Conference on Human
  Factors in Computing Systems}}. \bibinfo{pages}{599--606}.
\newblock
\urldef\tempurl%
\url{https://doi.org/10.1145/985692.985768}
\showDOI{\tempurl}


\bibitem[\protect\citeauthoryear{Cowan}{Cowan}{2011}]%
        {cowan2011lightweight}
\bibfield{author}{\bibinfo{person}{Lisa~G Cowan}.}
  \bibinfo{year}{2011}\natexlab{}.
\newblock \emph{\bibinfo{title}{Lightweight social communication using visual
  media and mobile phones}}.
\newblock \bibinfo{thesistype}{Ph.D. Dissertation}. \bibinfo{school}{UC San
  Diego}.
\newblock


\bibitem[\protect\citeauthoryear{Curran, Gordon, Lin, Sridhar, and
  Chuang}{Curran et~al\mbox{.}}{2019}]%
        {curran2019empathy}
\bibfield{author}{\bibinfo{person}{Max~T. Curran},
  \bibinfo{person}{Jeremy~Raboff Gordon}, \bibinfo{person}{Lily Lin},
  \bibinfo{person}{Priyashri~Kamlesh Sridhar}, {and} \bibinfo{person}{John
  Chuang}.} \bibinfo{year}{2019}\natexlab{}.
\newblock \showarticletitle{Understanding Digitally-Mediated Empathy: An
  Exploration of Visual, Narrative, and Biosensory Informational Cues}. In
  \bibinfo{booktitle}{\emph{Proceedings of the 2019 ACM Conference on Human
  Factors in Computing Systems}}. ACM, \bibinfo{pages}{614:1--614:13}.
\newblock
\urldef\tempurl%
\url{https://doi.org/10.1145/3290605.3300844}
\showDOI{\tempurl}


\bibitem[\protect\citeauthoryear{Dey and de~Guzman}{Dey and de~Guzman}{2006}]%
        {dey2006awareness}
\bibfield{author}{\bibinfo{person}{Anind~K Dey} {and} \bibinfo{person}{Ed de
  Guzman}.} \bibinfo{year}{2006}\natexlab{}.
\newblock \showarticletitle{From awareness to connectedness: the design and
  deployment of presence displays}. In \bibinfo{booktitle}{\emph{Proceedings of
  the SIGCHI Conference on Human Factors in Computing Systems}}. ACM,
  \bibinfo{pages}{899--908}.
\newblock
\urldef\tempurl%
\url{https://doi.org/10.1145/1124772.1124905}
\showDOI{\tempurl}


\bibitem[\protect\citeauthoryear{Egger, Ley, and Hanke}{Egger
  et~al\mbox{.}}{2019}]%
        {egger2019emotion}
\bibfield{author}{\bibinfo{person}{Maria Egger}, \bibinfo{person}{Matthias
  Ley}, {and} \bibinfo{person}{Sten Hanke}.} \bibinfo{year}{2019}\natexlab{}.
\newblock \showarticletitle{Emotion recognition from physiological signal
  analysis: A review}.
\newblock \bibinfo{journal}{\emph{Electronic Notes in Theoretical Computer
  Science}}  \bibinfo{volume}{343} (\bibinfo{year}{2019}),
  \bibinfo{pages}{35--55}.
\newblock
\urldef\tempurl%
\url{https://doi.org/10.1016/j.entcs.2019.04.009}
\showDOI{\tempurl}


\bibitem[\protect\citeauthoryear{Eichhorn, Wettach, and Hornecker}{Eichhorn
  et~al\mbox{.}}{2008}]%
        {eichhorn2008stroking}
\bibfield{author}{\bibinfo{person}{Elisabeth Eichhorn}, \bibinfo{person}{Reto
  Wettach}, {and} \bibinfo{person}{Eva Hornecker}.}
  \bibinfo{year}{2008}\natexlab{}.
\newblock \showarticletitle{A stroking device for spatially separated couples}.
  In \bibinfo{booktitle}{\emph{Proceedings of the 10th international conference
  on Human computer interaction with mobile devices and services}}. ACM,
  \bibinfo{pages}{303--306}.
\newblock
\urldef\tempurl%
\url{https://doi.org/10.1145/1409240.1409274}
\showDOI{\tempurl}


\bibitem[\protect\citeauthoryear{Griggio, Mcgrenere, and Mackay}{Griggio
  et~al\mbox{.}}{2019a}]%
        {griggio2019customizations}
\bibfield{author}{\bibinfo{person}{Carla~F Griggio}, \bibinfo{person}{Joanna
  Mcgrenere}, {and} \bibinfo{person}{Wendy~E Mackay}.}
  \bibinfo{year}{2019}\natexlab{a}.
\newblock \showarticletitle{Customizations and Expression Breakdowns in
  Ecosystems of Communication Apps}.
\newblock \bibinfo{journal}{\emph{Proceedings of the ACM on Human-Computer
  Interaction}} \bibinfo{volume}{3}, \bibinfo{number}{CSCW}
  (\bibinfo{year}{2019}), \bibinfo{pages}{1--26}.
\newblock
\urldef\tempurl%
\url{https://doi.org/10.1145/3359128}
\showDOI{\tempurl}


\bibitem[\protect\citeauthoryear{Griggio, Nouwens, Mcgrenere, and
  Mackay}{Griggio et~al\mbox{.}}{2019b}]%
        {griggio2019augmenting}
\bibfield{author}{\bibinfo{person}{Carla~F Griggio}, \bibinfo{person}{Midas
  Nouwens}, \bibinfo{person}{Joanna Mcgrenere}, {and} \bibinfo{person}{Wendy~E
  Mackay}.} \bibinfo{year}{2019}\natexlab{b}.
\newblock \showarticletitle{Augmenting Couples' Communication with Lifelines:
  Shared Timelines of Mixed Contextual Information}. In
  \bibinfo{booktitle}{\emph{Proceedings of the 2019 CHI Conference on Human
  Factors in Computing Systems}}. ACM, \bibinfo{pages}{623}.
\newblock
\urldef\tempurl%
\url{https://doi.org/10.1145/3290605.3300853}
\showDOI{\tempurl}


\bibitem[\protect\citeauthoryear{Haans and IJsselsteijn}{Haans and
  IJsselsteijn}{2006}]%
        {haans2006mediated}
\bibfield{author}{\bibinfo{person}{Antal Haans} {and} \bibinfo{person}{Wijnand
  IJsselsteijn}.} \bibinfo{year}{2006}\natexlab{}.
\newblock \showarticletitle{Mediated social touch: a review of current research
  and future directions}.
\newblock \bibinfo{journal}{\emph{Virtual Reality}} \bibinfo{volume}{9},
  \bibinfo{number}{2-3} (\bibinfo{year}{2006}), \bibinfo{pages}{149--159}.
\newblock
\urldef\tempurl%
\url{https://doi.org/10.1007/s10055-005-0014-2}
\showDOI{\tempurl}


\bibitem[\protect\citeauthoryear{Hassenzahl, Heidecker, Eckoldt, Diefenbach,
  and Hillmann}{Hassenzahl et~al\mbox{.}}{2012}]%
        {hassenzahl2012all}
\bibfield{author}{\bibinfo{person}{Marc Hassenzahl}, \bibinfo{person}{Stephanie
  Heidecker}, \bibinfo{person}{Kai Eckoldt}, \bibinfo{person}{Sarah
  Diefenbach}, {and} \bibinfo{person}{Uwe Hillmann}.}
  \bibinfo{year}{2012}\natexlab{}.
\newblock \showarticletitle{All you need is love: Current strategies of
  mediating intimate relationships through technology}.
\newblock \bibinfo{journal}{\emph{ACM Transactions on Computer-Human
  Interaction (TOCHI)}} \bibinfo{volume}{19}, \bibinfo{number}{4}
  (\bibinfo{year}{2012}), \bibinfo{pages}{1--19}.
\newblock
\urldef\tempurl%
\url{https://doi.org/10.1145/2395131.2395137}
\showDOI{\tempurl}


\bibitem[\protect\citeauthoryear{Hassib, Buschek, Wozniak, and Alt}{Hassib
  et~al\mbox{.}}{2017a}]%
        {hassib2017heartchat}
\bibfield{author}{\bibinfo{person}{Mariam Hassib}, \bibinfo{person}{Daniel
  Buschek}, \bibinfo{person}{PawelW~W Wozniak}, {and} \bibinfo{person}{Florian
  Alt}.} \bibinfo{year}{2017}\natexlab{a}.
\newblock \showarticletitle{HeartChat: Heart Rate Augmented Mobile Chat to
  Support Empathy and Awareness}. In \bibinfo{booktitle}{\emph{Proceedings of
  the 2017 CHI Conference on Human Factors in Computing Systems}}. ACM,
  \bibinfo{pages}{2239--2251}.
\newblock
\urldef\tempurl%
\url{https://doi.org/10.1145/3025453.3025758}
\showDOI{\tempurl}


\bibitem[\protect\citeauthoryear{Hassib, Schneegass, Eiglsperger, Henze,
  Schmidt, and Alt}{Hassib et~al\mbox{.}}{2017b}]%
        {hassib2017engagemeter}
\bibfield{author}{\bibinfo{person}{Mariam Hassib}, \bibinfo{person}{Stefan
  Schneegass}, \bibinfo{person}{Philipp Eiglsperger}, \bibinfo{person}{Niels
  Henze}, \bibinfo{person}{Albrecht Schmidt}, {and} \bibinfo{person}{Florian
  Alt}.} \bibinfo{year}{2017}\natexlab{b}.
\newblock \showarticletitle{EngageMeter: A System for Implicit Audience
  Engagement Sensing Using Electroencephalography}. In
  \bibinfo{booktitle}{\emph{Proceedings of the 2017 CHI Conference on Human
  Factors in Computing Systems}}. ACM, \bibinfo{pages}{5114--5119}.
\newblock
\urldef\tempurl%
\url{https://doi.org/10.1145/3025453.3025669}
\showDOI{\tempurl}


\bibitem[\protect\citeauthoryear{Hollis, Pekurovsky, Wu, and Whittaker}{Hollis
  et~al\mbox{.}}{2018}]%
        {hollis2018being}
\bibfield{author}{\bibinfo{person}{Victoria Hollis}, \bibinfo{person}{Alon
  Pekurovsky}, \bibinfo{person}{Eunika Wu}, {and} \bibinfo{person}{Steve
  Whittaker}.} \bibinfo{year}{2018}\natexlab{}.
\newblock \showarticletitle{On being told how we feel: how algorithmic sensor
  feedback influences emotion perception}.
\newblock \bibinfo{journal}{\emph{Proceedings of the ACM on Interactive,
  Mobile, Wearable and Ubiquitous Technologies}} \bibinfo{volume}{2},
  \bibinfo{number}{3} (\bibinfo{year}{2018}), \bibinfo{pages}{1--31}.
\newblock
\urldef\tempurl%
\url{https://doi.org/10.1145/3264924}
\showDOI{\tempurl}


\bibitem[\protect\citeauthoryear{Howell, Chuang, De~Kosnik, Niemeyer, and
  Ryokai}{Howell et~al\mbox{.}}{2018a}]%
        {howell2018emotional}
\bibfield{author}{\bibinfo{person}{Noura Howell}, \bibinfo{person}{John
  Chuang}, \bibinfo{person}{Abigail De~Kosnik}, \bibinfo{person}{Greg
  Niemeyer}, {and} \bibinfo{person}{Kimiko Ryokai}.}
  \bibinfo{year}{2018}\natexlab{a}.
\newblock \showarticletitle{Emotional Biosensing: Exploring Critical
  Alternatives}.
\newblock \bibinfo{journal}{\emph{Proceedings of the ACM on Human-Computer
  Interaction}} \bibinfo{volume}{2}, \bibinfo{number}{CSCW}
  (\bibinfo{year}{2018}), \bibinfo{pages}{69}.
\newblock
\urldef\tempurl%
\url{https://doi.org/10.1145/3274338}
\showDOI{\tempurl}


\bibitem[\protect\citeauthoryear{Howell, Devendorf, Tian, Vega~Galvez, Gong,
  Poupyrev, Paulos, and Ryokai}{Howell et~al\mbox{.}}{2016}]%
        {howell2016biosignals}
\bibfield{author}{\bibinfo{person}{Noura Howell}, \bibinfo{person}{Laura
  Devendorf}, \bibinfo{person}{Rundong~Kevin Tian}, \bibinfo{person}{Tom{\'a}s
  Vega~Galvez}, \bibinfo{person}{Nan-Wei Gong}, \bibinfo{person}{Ivan
  Poupyrev}, \bibinfo{person}{Eric Paulos}, {and} \bibinfo{person}{Kimiko
  Ryokai}.} \bibinfo{year}{2016}\natexlab{}.
\newblock \showarticletitle{Biosignals as Social Cues: Ambiguity and Emotional
  Interpretation in Social Displays of Skin Conductance}. In
  \bibinfo{booktitle}{\emph{Proceedings of the 2016 ACM Conference on Designing
  Interactive Systems}}. ACM, \bibinfo{pages}{865--870}.
\newblock
\urldef\tempurl%
\url{https://doi.org/10.1145/2901790.2901850}
\showDOI{\tempurl}


\bibitem[\protect\citeauthoryear{Howell, Devendorf, Vega~G{\'a}lvez, Tian, and
  Ryokai}{Howell et~al\mbox{.}}{2018b}]%
        {howell2018tensions}
\bibfield{author}{\bibinfo{person}{Noura Howell}, \bibinfo{person}{Laura
  Devendorf}, \bibinfo{person}{Tom{\'a}s~Alfonso Vega~G{\'a}lvez},
  \bibinfo{person}{Rundong Tian}, {and} \bibinfo{person}{Kimiko Ryokai}.}
  \bibinfo{year}{2018}\natexlab{b}.
\newblock \showarticletitle{Tensions of Data-Driven Reflection: A Case Study of
  Real-Time Emotional Biosensing}. In \bibinfo{booktitle}{\emph{Proceedings of
  the 2018 CHI Conference on Human Factors in Computing Systems}}. ACM,
  \bibinfo{pages}{431}.
\newblock
\urldef\tempurl%
\url{https://doi.org/10.1145/3173574.3174005}
\showDOI{\tempurl}


\bibitem[\protect\citeauthoryear{Howell, Niemeyer, and Ryokai}{Howell
  et~al\mbox{.}}{2019}]%
        {howell2019bench}
\bibfield{author}{\bibinfo{person}{Noura Howell}, \bibinfo{person}{Greg
  Niemeyer}, {and} \bibinfo{person}{Kimiko Ryokai}.}
  \bibinfo{year}{2019}\natexlab{}.
\newblock \showarticletitle{Life-Affirming Biosensing in Public: Sounding
  Heartbeats on a Red Bench}. In \bibinfo{booktitle}{\emph{Proceedings of the
  2019 CHI Conference on Human Factors in Computing Systems}}
  \emph{(\bibinfo{series}{CHI '19})}. \bibinfo{publisher}{ACM},
  \bibinfo{address}{New York, NY, USA}, Article \bibinfo{articleno}{680},
  \bibinfo{numpages}{16}~pages.
\newblock
\showISBNx{978-1-4503-5970-2}
\urldef\tempurl%
\url{https://doi.org/10.1145/3290605.3300910}
\showDOI{\tempurl}


\bibitem[\protect\citeauthoryear{IJsselsteijn, van Baren, and van
  Lanen}{IJsselsteijn et~al\mbox{.}}{2003}]%
        {ijsselsteijn2003staying}
\bibfield{author}{\bibinfo{person}{Wijnand IJsselsteijn}, \bibinfo{person}{Joy
  van Baren}, {and} \bibinfo{person}{Froukje van Lanen}.}
  \bibinfo{year}{2003}\natexlab{}.
\newblock \showarticletitle{Staying in touch: Social presence and connectedness
  through synchronous and asynchronous communication media}.
\newblock \bibinfo{journal}{\emph{Human-Computer Interaction: Theory and
  Practice (Part II)}} \bibinfo{volume}{2}, \bibinfo{number}{924}
  (\bibinfo{year}{2003}), \bibinfo{pages}{e928}.
\newblock


\bibitem[\protect\citeauthoryear{Jakesch, French, Ma, Hancock, and
  Naaman}{Jakesch et~al\mbox{.}}{2019}]%
        {jakesch2019ai}
\bibfield{author}{\bibinfo{person}{Maurice Jakesch}, \bibinfo{person}{Megan
  French}, \bibinfo{person}{Xiao Ma}, \bibinfo{person}{Jeffrey~T Hancock},
  {and} \bibinfo{person}{Mor Naaman}.} \bibinfo{year}{2019}\natexlab{}.
\newblock \showarticletitle{AI-mediated communication: How the perception that
  profile text was written by AI affects trustworthiness}. In
  \bibinfo{booktitle}{\emph{Proceedings of the 2019 CHI Conference on Human
  Factors in Computing Systems}}. \bibinfo{pages}{1--13}.
\newblock
\urldef\tempurl%
\url{https://doi.org/10.1145/3290605.3300469}
\showDOI{\tempurl}


\bibitem[\protect\citeauthoryear{Janssen, Bailenson, IJsselsteijn, and
  Westerink}{Janssen et~al\mbox{.}}{2010}]%
        {janssen2010intimate}
\bibfield{author}{\bibinfo{person}{Joris~H Janssen}, \bibinfo{person}{Jeremy~N
  Bailenson}, \bibinfo{person}{Wijnand~A IJsselsteijn}, {and}
  \bibinfo{person}{Joyce~HDM Westerink}.} \bibinfo{year}{2010}\natexlab{}.
\newblock \showarticletitle{Intimate Heartbeats: Opportunities for Affective
  Communication Technology}.
\newblock \bibinfo{journal}{\emph{IEEE Transactions on Affective Computing}}
  \bibinfo{number}{2} (\bibinfo{year}{2010}), \bibinfo{pages}{72--80}.
\newblock
\urldef\tempurl%
\url{https://doi.org/10.1109/T-AFFC.2010.13}
\showDOI{\tempurl}


\bibitem[\protect\citeauthoryear{Jones and Bodie}{Jones and Bodie}{2014}]%
        {jones201416}
\bibfield{author}{\bibinfo{person}{Susanne~M Jones} {and}
  \bibinfo{person}{Graham~D Bodie}.} \bibinfo{year}{2014}\natexlab{}.
\newblock \showarticletitle{Supportive communication}.
\newblock \bibinfo{journal}{\emph{Interpersonal communication}}
  \bibinfo{volume}{6} (\bibinfo{year}{2014}), \bibinfo{pages}{371}.
\newblock
\urldef\tempurl%
\url{https://doi.org/10.1515/9783110276794.371}
\showDOI{\tempurl}


\bibitem[\protect\citeauthoryear{Kaye}{Kaye}{2006}]%
        {kaye2006just}
\bibfield{author}{\bibinfo{person}{Joseph~`Jofish' Kaye}.}
  \bibinfo{year}{2006}\natexlab{}.
\newblock \showarticletitle{I just clicked to say I love you: rich evaluations
  of minimal communication}. In \bibinfo{booktitle}{\emph{CHI'06 Extended
  Abstracts on Human Factors in Computing Systems}}. \bibinfo{pages}{363--368}.
\newblock
\urldef\tempurl%
\url{https://doi.org/10.1145/1125451.1125530}
\showDOI{\tempurl}


\bibitem[\protect\citeauthoryear{Kaye, Levitt, Nevins, Golden, and
  Schmidt}{Kaye et~al\mbox{.}}{2005}]%
        {kaye2005communicating}
\bibfield{author}{\bibinfo{person}{Joseph~`Jofish' Kaye},
  \bibinfo{person}{Mariah~K Levitt}, \bibinfo{person}{Jeffrey Nevins},
  \bibinfo{person}{Jessica Golden}, {and} \bibinfo{person}{Vanessa Schmidt}.}
  \bibinfo{year}{2005}\natexlab{}.
\newblock \showarticletitle{Communicating intimacy one bit at a time}. In
  \bibinfo{booktitle}{\emph{CHI'05 Extended Abstracts on Human Factors in
  Computing Systems}}. ACM, \bibinfo{pages}{1529--1532}.
\newblock
\urldef\tempurl%
\url{https://doi.org/10.1145/1056808.1056958}
\showDOI{\tempurl}


\bibitem[\protect\citeauthoryear{Kelly, Gooch, Patil, and Watts}{Kelly
  et~al\mbox{.}}{2017}]%
        {kelly2017demanding}
\bibfield{author}{\bibinfo{person}{Ryan Kelly}, \bibinfo{person}{Daniel Gooch},
  \bibinfo{person}{Bhagyashree Patil}, {and} \bibinfo{person}{Leon Watts}.}
  \bibinfo{year}{2017}\natexlab{}.
\newblock \showarticletitle{Demanding by design: Supporting effortful
  communication practices in close personal relationships}. In
  \bibinfo{booktitle}{\emph{Proceedings of the 2017 ACM Conference on Computer
  Supported Cooperative Work and Social Computing}}. \bibinfo{pages}{70--83}.
\newblock
\urldef\tempurl%
\url{https://doi.org/10.1145/2998181.2998184}
\showDOI{\tempurl}


\bibitem[\protect\citeauthoryear{Kelly, Gooch, and Watts}{Kelly
  et~al\mbox{.}}{2018}]%
        {kelly2018s}
\bibfield{author}{\bibinfo{person}{Ryan Kelly}, \bibinfo{person}{Daniel Gooch},
  {and} \bibinfo{person}{Leon Watts}.} \bibinfo{year}{2018}\natexlab{}.
\newblock \showarticletitle{`It's More Like a Letter' An Exploration of
  Mediated Conversational Effort in Message Builder}.
\newblock \bibinfo{journal}{\emph{Proceedings of the ACM on Human-Computer
  Interaction}} \bibinfo{volume}{2}, \bibinfo{number}{CSCW}
  (\bibinfo{year}{2018}), \bibinfo{pages}{1--23}.
\newblock
\urldef\tempurl%
\url{https://doi.org/10.1145/3274356}
\showDOI{\tempurl}


\bibitem[\protect\citeauthoryear{Kiesler, Siegel, and McGuire}{Kiesler
  et~al\mbox{.}}{1984}]%
        {kiesler1984social}
\bibfield{author}{\bibinfo{person}{Sara Kiesler}, \bibinfo{person}{Jane
  Siegel}, {and} \bibinfo{person}{Timothy~W McGuire}.}
  \bibinfo{year}{1984}\natexlab{}.
\newblock \showarticletitle{Social Psychological Aspects of Computer-Mediated
  Communication.}
\newblock \bibinfo{journal}{\emph{American Psychologist}} \bibinfo{volume}{39},
  \bibinfo{number}{10} (\bibinfo{year}{1984}), \bibinfo{pages}{1123}.
\newblock
\urldef\tempurl%
\url{https://doi.org/10.1037/0003-066X.39.10.1123}
\showDOI{\tempurl}


\bibitem[\protect\citeauthoryear{Kim}{Kim}{2014}]%
        {bedtimes}
\bibfield{author}{\bibinfo{person}{Susanna Kim}.}
  \bibinfo{year}{2014}\natexlab{}.
\newblock \bibinfo{title}{Cities' Bedtimes Revealed in One Map}.
\newblock
\newblock
\urldef\tempurl%
\url{https://abcnews.go.com/Business/map-claims-show-people-sleep/story?id=26042978}
\showURL{%
\tempurl}


\bibitem[\protect\citeauthoryear{Kirilenko and Stepchenkova}{Kirilenko and
  Stepchenkova}{2016}]%
        {kirilenko2016inter}
\bibfield{author}{\bibinfo{person}{Andrei~P Kirilenko} {and}
  \bibinfo{person}{Svetlana Stepchenkova}.} \bibinfo{year}{2016}\natexlab{}.
\newblock \showarticletitle{Inter-coder agreement in one-to-many
  classification: fuzzy kappa}.
\newblock \bibinfo{journal}{\emph{PloS one}} \bibinfo{volume}{11},
  \bibinfo{number}{3} (\bibinfo{year}{2016}), \bibinfo{pages}{e0149787}.
\newblock
\urldef\tempurl%
\url{https://doi.org/10.1371/journal.pone.0149787}
\showDOI{\tempurl}


\bibitem[\protect\citeauthoryear{Koeze and Popper}{Koeze and Popper}{2020}]%
        {nytimes_covid}
\bibfield{author}{\bibinfo{person}{Ella Koeze} {and} \bibinfo{person}{Nathaniel
  Popper}.} \bibinfo{year}{2020}\natexlab{}.
\newblock \bibinfo{title}{The Virus Changed the Way We Internet}.
\newblock
\newblock
\urldef\tempurl%
\url{https://www.nytimes.com/interactive/2020/04/07/technology/coronavirus-internet-use.html}
\showURL{%
\tempurl}


\bibitem[\protect\citeauthoryear{Kowalski, Loehmann, and Hausen}{Kowalski
  et~al\mbox{.}}{2013}]%
        {kowalski2013cubble}
\bibfield{author}{\bibinfo{person}{Robert Kowalski}, \bibinfo{person}{Sebastian
  Loehmann}, {and} \bibinfo{person}{Doris Hausen}.}
  \bibinfo{year}{2013}\natexlab{}.
\newblock \showarticletitle{Cubble: A multi-device hybrid approach supporting
  communication in long-distance relationships}. In
  \bibinfo{booktitle}{\emph{Proceedings of the 7th International Conference on
  Tangible, Embedded and Embodied Interaction}}. ACM,
  \bibinfo{pages}{201--204}.
\newblock
\urldef\tempurl%
\url{https://doi.org/10.1145/2460625.2460656}
\showDOI{\tempurl}


\bibitem[\protect\citeauthoryear{Kr{\"a}uchi and Wirz-Justice}{Kr{\"a}uchi and
  Wirz-Justice}{2001}]%
        {krauchi2001circadian}
\bibfield{author}{\bibinfo{person}{Kurt Kr{\"a}uchi} {and}
  \bibinfo{person}{Anna Wirz-Justice}.} \bibinfo{year}{2001}\natexlab{}.
\newblock \showarticletitle{Circadian clues to sleep onset mechanisms}.
\newblock \bibinfo{journal}{\emph{Neuropsychopharmacology}}
  \bibinfo{volume}{25}, \bibinfo{number}{1} (\bibinfo{year}{2001}),
  \bibinfo{pages}{S92--S96}.
\newblock
\urldef\tempurl%
\url{https://doi.org/10.1016/S0893-133X(01)00315-3}
\showDOI{\tempurl}


\bibitem[\protect\citeauthoryear{Larson}{Larson}{2002}]%
        {larson2002dinner}
\bibfield{author}{\bibinfo{person}{Ronald~B Larson}.}
  \bibinfo{year}{2002}\natexlab{}.
\newblock \showarticletitle{When is dinner?}
\newblock \bibinfo{journal}{\emph{Journal of Food Distribution Research}}
  \bibinfo{volume}{33}, \bibinfo{number}{856-2016-57369}
  (\bibinfo{year}{2002}), \bibinfo{pages}{38--45}.
\newblock
\urldef\tempurl%
\url{https://doi.org/10.22004/ag.econ.26834}
\showDOI{\tempurl}


\bibitem[\protect\citeauthoryear{Laurenceau, Barrett, and
  Pietromonaco}{Laurenceau et~al\mbox{.}}{1998}]%
        {laurenceau1998intimacy}
\bibfield{author}{\bibinfo{person}{Jean-Philippe Laurenceau},
  \bibinfo{person}{Lisa~Feldman Barrett}, {and} \bibinfo{person}{Paula~R
  Pietromonaco}.} \bibinfo{year}{1998}\natexlab{}.
\newblock \showarticletitle{Intimacy as an interpersonal process: The
  importance of self-disclosure, partner disclosure, and perceived partner
  responsiveness in interpersonal exchanges.}
\newblock \bibinfo{journal}{\emph{Journal of personality and social
  psychology}} \bibinfo{volume}{74}, \bibinfo{number}{5}
  (\bibinfo{year}{1998}), \bibinfo{pages}{1238}.
\newblock
\urldef\tempurl%
\url{https://doi.org/10.1037//0022-3514.74.5.1238}
\showDOI{\tempurl}


\bibitem[\protect\citeauthoryear{Liu, Dabbish, and Kaufman}{Liu
  et~al\mbox{.}}{2017a}]%
        {liu2017expressive}
\bibfield{author}{\bibinfo{person}{Fannie Liu}, \bibinfo{person}{Laura
  Dabbish}, {and} \bibinfo{person}{Geoff Kaufman}.}
  \bibinfo{year}{2017}\natexlab{a}.
\newblock \showarticletitle{Can Biosignals be Expressive? How Visualizations
  Affect Impression Formation from Shared Brain Activity}.
\newblock \bibinfo{journal}{\emph{Proceedings of the ACM on Human-Computer
  Interaction}} \bibinfo{volume}{1}, \bibinfo{number}{CSCW}
  (\bibinfo{year}{2017}), \bibinfo{pages}{71:1--71:21}.
\newblock
\urldef\tempurl%
\url{https://doi.org/10.1145/3134706}
\showDOI{\tempurl}


\bibitem[\protect\citeauthoryear{Liu, Dabbish, and Kaufman}{Liu
  et~al\mbox{.}}{2017b}]%
        {liu2017supporting}
\bibfield{author}{\bibinfo{person}{Fannie Liu}, \bibinfo{person}{Laura
  Dabbish}, {and} \bibinfo{person}{Geoff Kaufman}.}
  \bibinfo{year}{2017}\natexlab{b}.
\newblock \showarticletitle{Supporting Social Interactions with an Expressive
  Heart Rate Sharing Application}.
\newblock \bibinfo{journal}{\emph{Proceedings of the ACM on Interactive,
  Mobile, Wearable and Ubiquitous Technologies}} \bibinfo{volume}{1},
  \bibinfo{number}{3} (\bibinfo{year}{2017}), \bibinfo{pages}{77:1--77:26}.
\newblock
\urldef\tempurl%
\url{https://doi.org/10.1145/3130943}
\showDOI{\tempurl}


\bibitem[\protect\citeauthoryear{Liu, Esparza, Pavlovskaia, Kaufman, Dabbish,
  and Monroy-Hern\'{a}ndez}{Liu et~al\mbox{.}}{2019a}]%
        {liu2019animo}
\bibfield{author}{\bibinfo{person}{Fannie Liu}, \bibinfo{person}{Mario
  Esparza}, \bibinfo{person}{Maria Pavlovskaia}, \bibinfo{person}{Geoff
  Kaufman}, \bibinfo{person}{Laura Dabbish}, {and} \bibinfo{person}{Andr{\'e}s
  Monroy-Hern\'{a}ndez}.} \bibinfo{year}{2019}\natexlab{a}.
\newblock \showarticletitle{Animo: Sharing Biosignals on a Smartwatch for
  Lightweight Social Connection}.
\newblock \bibinfo{journal}{\emph{Proceedings of the ACM on Interactive,
  Mobile, Wearable and Ubiquitous Technologies}} \bibinfo{volume}{3},
  \bibinfo{number}{1} (\bibinfo{year}{2019}), \bibinfo{pages}{18:1--18:19}.
\newblock
\urldef\tempurl%
\url{https://doi.org/10.1145/3314405}
\showDOI{\tempurl}


\bibitem[\protect\citeauthoryear{Liu, Kaufman, and Dabbish}{Liu
  et~al\mbox{.}}{2019b}]%
        {liu2019empathy}
\bibfield{author}{\bibinfo{person}{Fannie Liu}, \bibinfo{person}{Geoff
  Kaufman}, {and} \bibinfo{person}{Laura Dabbish}.}
  \bibinfo{year}{2019}\natexlab{b}.
\newblock \showarticletitle{The Effect of Expressive Biosignals on Empathy and
  Closeness for a Stigmatized Group Member}.
\newblock \bibinfo{journal}{\emph{Proceedings of the ACM on Human-Computer
  Interaction}} \bibinfo{volume}{3}, \bibinfo{number}{CSCW}
  (\bibinfo{year}{2019}), \bibinfo{pages}{201:1--201:17}.
\newblock
\urldef\tempurl%
\url{https://doi.org/10.1145/3359303}
\showDOI{\tempurl}


\bibitem[\protect\citeauthoryear{Lo}{Lo}{2008}]%
        {lo2008nonverbal}
\bibfield{author}{\bibinfo{person}{Shao-Kang Lo}.}
  \bibinfo{year}{2008}\natexlab{}.
\newblock \showarticletitle{The nonverbal communication functions of emoticons
  in computer-mediated communication}.
\newblock \bibinfo{journal}{\emph{CyberPsychology \& Behavior}}
  \bibinfo{volume}{11}, \bibinfo{number}{5} (\bibinfo{year}{2008}),
  \bibinfo{pages}{595--597}.
\newblock
\urldef\tempurl%
\url{https://doi.org/10.1089/cpb.2007.0132}
\showDOI{\tempurl}


\bibitem[\protect\citeauthoryear{MacGeorge, Feng, and Burleson}{MacGeorge
  et~al\mbox{.}}{2011}]%
        {macgeorge2011supportive}
\bibfield{author}{\bibinfo{person}{Erina~L. MacGeorge}, \bibinfo{person}{Bo
  Feng}, {and} \bibinfo{person}{Brant~R. Burleson}.}
  \bibinfo{year}{2011}\natexlab{}.
\newblock \showarticletitle{Supportive Communication}.
\newblock In \bibinfo{booktitle}{\emph{Handbook of Interpersonal
  Communication}}, \bibfield{editor}{\bibinfo{person}{Mark~L. Knapp} {and}
  \bibinfo{person}{John~A. Daly}} (Eds.). \bibinfo{publisher}{SAGE},
  \bibinfo{address}{Thousand Oaks, CA}, \bibinfo{pages}{317--354}.
\newblock


\bibitem[\protect\citeauthoryear{Merrill and Cheshire}{Merrill and
  Cheshire}{2017}]%
        {merrill2017trust}
\bibfield{author}{\bibinfo{person}{Nick Merrill} {and} \bibinfo{person}{Coye
  Cheshire}.} \bibinfo{year}{2017}\natexlab{}.
\newblock \showarticletitle{Trust Your Heart: Assessing Cooperation and Trust
  with Biosignals in Computer-Mediated Interactions}. In
  \bibinfo{booktitle}{\emph{Proceedings of the 2017 ACM Conference on Computer
  Supported Cooperative Work and Social Computing}}. ACM,
  \bibinfo{pages}{2--12}.
\newblock
\urldef\tempurl%
\url{https://doi.org/10.1145/2998181.2998286}
\showDOI{\tempurl}


\bibitem[\protect\citeauthoryear{Merrill, Chuang, and Cheshire}{Merrill
  et~al\mbox{.}}{2019}]%
        {merrill2019sensing}
\bibfield{author}{\bibinfo{person}{Nick Merrill}, \bibinfo{person}{John
  Chuang}, {and} \bibinfo{person}{Coye Cheshire}.}
  \bibinfo{year}{2019}\natexlab{}.
\newblock \showarticletitle{Sensing is Believing: What People Think Biosensors
  Can Reveal About Thoughts and Feelings}. In
  \bibinfo{booktitle}{\emph{Proceedings of the 2019 on Designing Interactive
  Systems Conference}}. \bibinfo{pages}{413--420}.
\newblock
\urldef\tempurl%
\url{https://doi.org/10.1145/3322276.3322286}
\showDOI{\tempurl}


\bibitem[\protect\citeauthoryear{Miller, Kluver, Thebault-Spieker, Terveen, and
  Hecht}{Miller et~al\mbox{.}}{2017}]%
        {miller2017understanding}
\bibfield{author}{\bibinfo{person}{Hannah Miller}, \bibinfo{person}{Daniel
  Kluver}, \bibinfo{person}{Jacob Thebault-Spieker}, \bibinfo{person}{Loren
  Terveen}, {and} \bibinfo{person}{Brent Hecht}.}
  \bibinfo{year}{2017}\natexlab{}.
\newblock \showarticletitle{Understanding emoji ambiguity in context: The role
  of text in emoji-related miscommunication}. In \bibinfo{booktitle}{\emph{11th
  International Conference on Web and Social Media, ICWSM 2017}}. AAAI Press.
\newblock


\bibitem[\protect\citeauthoryear{Miller, Thebault-Spieker, Chang, Johnson,
  Terveen, and Hecht}{Miller et~al\mbox{.}}{2016}]%
        {miller2016blissfully}
\bibfield{author}{\bibinfo{person}{Hannah~Jean Miller}, \bibinfo{person}{Jacob
  Thebault-Spieker}, \bibinfo{person}{Shuo Chang}, \bibinfo{person}{Isaac
  Johnson}, \bibinfo{person}{Loren Terveen}, {and} \bibinfo{person}{Brent
  Hecht}.} \bibinfo{year}{2016}\natexlab{}.
\newblock \showarticletitle{"Blissfully Happy" or "Ready to Fight": Varying
  Interpretations of Emoji}. In \bibinfo{booktitle}{\emph{Proceedings of the
  10th International Conference on Web and Social Media}}.
\newblock


\bibitem[\protect\citeauthoryear{Min and Nam}{Min and Nam}{2014}]%
        {min2014biosignal}
\bibfield{author}{\bibinfo{person}{Hyeryung~Christine Min} {and}
  \bibinfo{person}{Tek-Jin Nam}.} \bibinfo{year}{2014}\natexlab{}.
\newblock \showarticletitle{Biosignal sharing for affective connectedness}. In
  \bibinfo{booktitle}{\emph{CHI'14 Extended Abstracts on Human Factors in
  Computing Systems}}. ACM, \bibinfo{pages}{2191--2196}.
\newblock
\urldef\tempurl%
\url{https://doi.org/10.1145/2559206.2581345}
\showDOI{\tempurl}


\bibitem[\protect\citeauthoryear{Monroy-Hern{\'a}ndez, Hill, Gonzalez-Rivero,
  and boyd}{Monroy-Hern{\'a}ndez et~al\mbox{.}}{2011}]%
        {monroy2011computers}
\bibfield{author}{\bibinfo{person}{Andr{\'e}s Monroy-Hern{\'a}ndez},
  \bibinfo{person}{Benjamin~Mako Hill}, \bibinfo{person}{Jazmin
  Gonzalez-Rivero}, {and} \bibinfo{person}{danah boyd}.}
  \bibinfo{year}{2011}\natexlab{}.
\newblock \showarticletitle{Computers can't give credit: How automatic
  attribution falls short in an online remixing community}. In
  \bibinfo{booktitle}{\emph{Proceedings of the SIGCHI Conference on Human
  Factors in Computing Systems}}. \bibinfo{pages}{3421--3430}.
\newblock
\urldef\tempurl%
\url{https://doi.org/10.1145/1978942.1979452}
\showDOI{\tempurl}


\bibitem[\protect\citeauthoryear{Mueller, Vetere, Gibbs, Kjeldskov, Pedell, and
  Howard}{Mueller et~al\mbox{.}}{2005}]%
        {mueller2005hug}
\bibfield{author}{\bibinfo{person}{Florian~`Floyd' Mueller},
  \bibinfo{person}{Frank Vetere}, \bibinfo{person}{Martin~R Gibbs},
  \bibinfo{person}{Jesper Kjeldskov}, \bibinfo{person}{Sonja Pedell}, {and}
  \bibinfo{person}{Steve Howard}.} \bibinfo{year}{2005}\natexlab{}.
\newblock \showarticletitle{Hug over a distance}. In
  \bibinfo{booktitle}{\emph{CHI'05 Extended Abstracts on Human Factors in
  Computing Systems}}. \bibinfo{pages}{1673--1676}.
\newblock
\urldef\tempurl%
\url{https://doi.org/10.1145/1056808.1056994}
\showDOI{\tempurl}


\bibitem[\protect\citeauthoryear{Posner, Russell, and Peterson}{Posner
  et~al\mbox{.}}{2005}]%
        {posner2005circumplex}
\bibfield{author}{\bibinfo{person}{Jonathan Posner}, \bibinfo{person}{James~A
  Russell}, {and} \bibinfo{person}{Bradley~S Peterson}.}
  \bibinfo{year}{2005}\natexlab{}.
\newblock \showarticletitle{The circumplex model of affect: An integrative
  approach to affective neuroscience, cognitive development, and
  psychopathology}.
\newblock \bibinfo{journal}{\emph{Development and Psychopathology}}
  \bibinfo{volume}{17}, \bibinfo{number}{3} (\bibinfo{year}{2005}),
  \bibinfo{pages}{715--734}.
\newblock
\urldef\tempurl%
\url{https://doi.org/10.1017/S0954579405050340}
\showDOI{\tempurl}


\bibitem[\protect\citeauthoryear{Sauder, Johnston, Skulas-Ray, Campbell, and
  West}{Sauder et~al\mbox{.}}{2012}]%
        {sauder2012effect}
\bibfield{author}{\bibinfo{person}{Katherine~A Sauder},
  \bibinfo{person}{Elyse~R Johnston}, \bibinfo{person}{Ann~C Skulas-Ray},
  \bibinfo{person}{Tavis~S Campbell}, {and} \bibinfo{person}{Sheila~G West}.}
  \bibinfo{year}{2012}\natexlab{}.
\newblock \showarticletitle{Effect of meal content on heart rate variability
  and cardiovascular reactivity to mental stress}.
\newblock \bibinfo{journal}{\emph{Psychophysiology}} \bibinfo{volume}{49},
  \bibinfo{number}{4} (\bibinfo{year}{2012}), \bibinfo{pages}{470--477}.
\newblock
\urldef\tempurl%
\url{https://doi.org/10.1111/j.1469-8986.2011.01335.x}
\showDOI{\tempurl}


\bibitem[\protect\citeauthoryear{Semertzidis, Scary, Andres, Dwivedi, Kulwe,
  Zambetta, and Mueller}{Semertzidis et~al\mbox{.}}{2020}]%
        {semertzidis2020neo}
\bibfield{author}{\bibinfo{person}{Nathan Semertzidis},
  \bibinfo{person}{Michaela Scary}, \bibinfo{person}{Josh Andres},
  \bibinfo{person}{Brahmi Dwivedi}, \bibinfo{person}{Yutika~Chandrashekhar
  Kulwe}, \bibinfo{person}{Fabio Zambetta}, {and}
  \bibinfo{person}{Florian~Floyd Mueller}.} \bibinfo{year}{2020}\natexlab{}.
\newblock \showarticletitle{Neo-Noumena: Augmenting Emotion Communication}. In
  \bibinfo{booktitle}{\emph{Proceedings of the 2020 CHI Conference on Human
  Factors in Computing Systems}}. \bibinfo{pages}{1--13}.
\newblock
\urldef\tempurl%
\url{https://doi.org/10.1145/3313831.3376599}
\showDOI{\tempurl}


\bibitem[\protect\citeauthoryear{Seppala, Rossomando, and Doty}{Seppala
  et~al\mbox{.}}{2013}]%
        {seppala2013social}
\bibfield{author}{\bibinfo{person}{Emma Seppala}, \bibinfo{person}{Timothy
  Rossomando}, {and} \bibinfo{person}{James~R Doty}.}
  \bibinfo{year}{2013}\natexlab{}.
\newblock \showarticletitle{Social connection and compassion: Important
  predictors of health and well-being}.
\newblock \bibinfo{journal}{\emph{Social Research: An International Quarterly}}
  \bibinfo{volume}{80}, \bibinfo{number}{2} (\bibinfo{year}{2013}),
  \bibinfo{pages}{411--430}.
\newblock
\urldef\tempurl%
\url{https://doi.org/10.1353/sor.2013.0027}
\showDOI{\tempurl}


\bibitem[\protect\citeauthoryear{Slov{\'a}k, Janssen, and
  Fitzpatrick}{Slov{\'a}k et~al\mbox{.}}{2012}]%
        {slovak2012understanding}
\bibfield{author}{\bibinfo{person}{Petr Slov{\'a}k}, \bibinfo{person}{Joris
  Janssen}, {and} \bibinfo{person}{Geraldine Fitzpatrick}.}
  \bibinfo{year}{2012}\natexlab{}.
\newblock \showarticletitle{Understanding Heart Rate Sharing: Towards Unpacking
  Physiosocial Space}. In \bibinfo{booktitle}{\emph{Proceedings of the SIGCHI
  Conference on Human Factors in Computing Systems}}. ACM,
  \bibinfo{pages}{859--868}.
\newblock
\urldef\tempurl%
\url{https://doi.org/10.1145/2207676.2208526}
\showDOI{\tempurl}


\bibitem[\protect\citeauthoryear{Strauss and Corbin}{Strauss and
  Corbin}{1998}]%
        {strauss1998basics}
\bibfield{author}{\bibinfo{person}{A Strauss} {and} \bibinfo{person}{J
  Corbin}.} \bibinfo{year}{1998}\natexlab{}.
\newblock \showarticletitle{Basics of qualitative research techniques}.
\newblock  (\bibinfo{year}{1998}).
\newblock
\urldef\tempurl%
\url{https://doi.org/10.4135/9781452230153}
\showDOI{\tempurl}


\bibitem[\protect\citeauthoryear{Tang and Hew}{Tang and Hew}{2018}]%
        {tang2018emoticon}
\bibfield{author}{\bibinfo{person}{Ying Tang} {and} \bibinfo{person}{Khe~Foon
  Hew}.} \bibinfo{year}{2018}\natexlab{}.
\newblock \showarticletitle{Emoticon, emoji, and sticker use in
  computer-mediated communications: Understanding its communicative function,
  impact, user behavior, and motive}.
\newblock In \bibinfo{booktitle}{\emph{New Media for Educational Change}}.
  \bibinfo{publisher}{Springer}, \bibinfo{pages}{191--201}.
\newblock
\urldef\tempurl%
\url{https://doi.org/10.1007/978-981-10-8896-4_16}
\showDOI{\tempurl}


\bibitem[\protect\citeauthoryear{Tanis and Postmes}{Tanis and Postmes}{2003}]%
        {tanis2003social}
\bibfield{author}{\bibinfo{person}{Martin Tanis} {and} \bibinfo{person}{Tom
  Postmes}.} \bibinfo{year}{2003}\natexlab{}.
\newblock \showarticletitle{Social cues and impression formation in CMC}.
\newblock \bibinfo{journal}{\emph{Journal of Communication}}
  \bibinfo{volume}{53}, \bibinfo{number}{4} (\bibinfo{year}{2003}),
  \bibinfo{pages}{676--693}.
\newblock
\urldef\tempurl%
\url{https://doi.org/10.1111/j.1460-2466.2003.tb02917.x}
\showDOI{\tempurl}


\bibitem[\protect\citeauthoryear{Tian, Galery, Dulcinati, Molimpakis, and
  Sun}{Tian et~al\mbox{.}}{2017}]%
        {tian2017facebook}
\bibfield{author}{\bibinfo{person}{Ye Tian}, \bibinfo{person}{Thiago Galery},
  \bibinfo{person}{Giulio Dulcinati}, \bibinfo{person}{Emilia Molimpakis},
  {and} \bibinfo{person}{Chao Sun}.} \bibinfo{year}{2017}\natexlab{}.
\newblock \showarticletitle{Facebook sentiment: Reactions and emojis}. In
  \bibinfo{booktitle}{\emph{Proceedings of the Fifth International Workshop on
  Natural Language Processing for Social Media}}. \bibinfo{pages}{11--16}.
\newblock
\urldef\tempurl%
\url{https://doi.org/10.18653/v1/W17-1102}
\showDOI{\tempurl}


\bibitem[\protect\citeauthoryear{Uchida, Townsend, Rose~Markus, and
  Bergsieker}{Uchida et~al\mbox{.}}{2009}]%
        {uchida2009emotions}
\bibfield{author}{\bibinfo{person}{Yukiko Uchida}, \bibinfo{person}{Sarah~SM
  Townsend}, \bibinfo{person}{Hazel Rose~Markus}, {and}
  \bibinfo{person}{Hilary~B Bergsieker}.} \bibinfo{year}{2009}\natexlab{}.
\newblock \showarticletitle{Emotions as within or between people? Cultural
  variation in lay theories of emotion expression and inference}.
\newblock \bibinfo{journal}{\emph{Personality and social psychology bulletin}}
  \bibinfo{volume}{35}, \bibinfo{number}{11} (\bibinfo{year}{2009}),
  \bibinfo{pages}{1427--1439}.
\newblock
\urldef\tempurl%
\url{https://doi.org/10.1177/0146167209347322}
\showDOI{\tempurl}


\bibitem[\protect\citeauthoryear{Walther}{Walther}{2011}]%
        {walther2011theories}
\bibfield{author}{\bibinfo{person}{Joseph~B Walther}.}
  \bibinfo{year}{2011}\natexlab{}.
\newblock \showarticletitle{Theories of computer-mediated communication and
  interpersonal relations}.
\newblock \bibinfo{journal}{\emph{The handbook of interpersonal communication}}
   \bibinfo{volume}{4} (\bibinfo{year}{2011}), \bibinfo{pages}{443--479}.
\newblock


\bibitem[\protect\citeauthoryear{Werner, Wettach, and Hornecker}{Werner
  et~al\mbox{.}}{2008}]%
        {werner2008united}
\bibfield{author}{\bibinfo{person}{Julia Werner}, \bibinfo{person}{Reto
  Wettach}, {and} \bibinfo{person}{Eva Hornecker}.}
  \bibinfo{year}{2008}\natexlab{}.
\newblock \showarticletitle{United-pulse: Feeling your partner's pulse}. In
  \bibinfo{booktitle}{\emph{Proceedings of the 10th Conference on
  Human-Computer Interaction with Mobile Devices and Services}}. ACM,
  \bibinfo{pages}{535--538}.
\newblock
\urldef\tempurl%
\url{https://doi.org/10.1145/1409240.1409338}
\showDOI{\tempurl}


\bibitem[\protect\citeauthoryear{West and Turner}{West and Turner}{2018}]%
        {west2018introducing}
\bibfield{author}{\bibinfo{person}{Richard West} {and} \bibinfo{person}{Lynn~H
  Turner}.} \bibinfo{year}{2018}\natexlab{}.
\newblock \bibinfo{booktitle}{\emph{Introducing Communication Theory: Analysis
  And Application}}.
\newblock \bibinfo{publisher}{McGraw-Hill Education,}.
\newblock


\bibitem[\protect\citeauthoryear{Wiese, Kelley, Cranor, Dabbish, Hong, and
  Zimmerman}{Wiese et~al\mbox{.}}{2011}]%
        {wiese2011you}
\bibfield{author}{\bibinfo{person}{Jason Wiese}, \bibinfo{person}{Patrick~Gage
  Kelley}, \bibinfo{person}{Lorrie~Faith Cranor}, \bibinfo{person}{Laura
  Dabbish}, \bibinfo{person}{Jason~I Hong}, {and} \bibinfo{person}{John
  Zimmerman}.} \bibinfo{year}{2011}\natexlab{}.
\newblock \showarticletitle{Are you close with me? are you nearby?:
  investigating social groups, closeness, and willingness to share}. In
  \bibinfo{booktitle}{\emph{Proceedings of the 13th International Conference on
  Ubiquitous Computing}}. ACM, \bibinfo{pages}{197--206}.
\newblock
\urldef\tempurl%
\url{https://doi.org/10.1145/2030112.2030140}
\showDOI{\tempurl}


\bibitem[\protect\citeauthoryear{Wilson and Atkeson}{Wilson and
  Atkeson}{2003}]%
        {wilson2003narrator}
\bibfield{author}{\bibinfo{person}{Daniel~H Wilson} {and}
  \bibinfo{person}{Christopher Atkeson}.} \bibinfo{year}{2003}\natexlab{}.
\newblock \showarticletitle{The narrator: A daily activity summarizer using
  simple sensors in an instrumented environment}. In
  \bibinfo{booktitle}{\emph{The Fifth International Conference on Ubiquitous
  Computing 2003 Demonstrations}}.
\newblock


\bibitem[\protect\citeauthoryear{Wiseman and Gould}{Wiseman and Gould}{2018}]%
        {wiseman2018repurposing}
\bibfield{author}{\bibinfo{person}{Sarah Wiseman} {and}
  \bibinfo{person}{Sandy~JJ Gould}.} \bibinfo{year}{2018}\natexlab{}.
\newblock \showarticletitle{Repurposing Emoji for Personalised Communication:
  Why means ``I love you''}. In \bibinfo{booktitle}{\emph{Proceedings of the
  2018 CHI Conference on Human Factors in Computing Systems}}. ACM,
  \bibinfo{pages}{152}.
\newblock
\urldef\tempurl%
\url{https://doi.org/10.1145/3173574.3173726}
\showDOI{\tempurl}


\bibitem[\protect\citeauthoryear{Zammuner}{Zammuner}{2000}]%
        {zammuner2000men}
\bibfield{author}{\bibinfo{person}{Vanda~L Zammuner}.}
  \bibinfo{year}{2000}\natexlab{}.
\newblock \showarticletitle{Men's and women's lay theories of emotion}.
\newblock \bibinfo{journal}{\emph{Gender and emotion: Social psychological
  perspectives}} (\bibinfo{year}{2000}), \bibinfo{pages}{48--70}.
\newblock
\urldef\tempurl%
\url{https://doi.org/10.1017/CBO9780511628191.004}
\showDOI{\tempurl}


\end{thebibliography}



\section*{Supplementary material}

# Daily Survey Questions

1. Please describe one noteworthy experience you had with the app today (e.g.,whether you noticed anything new, learned a new function, used the app in a new way).

**Sending your State Otter**
STATE otters are the otters you see on your main app screen that you can send to your partner to "initiate" an interaction. REACT otters are the ones on the "Tap to React" screen that you can send to your partner to "respond" to their otter.

Our records indicate that you sent the following state otter animation today, around (time).
<GIF of the state animation sent>
Please answer the questions below about this animation.

1. How did you first notice this state otter animation?
   (a) "Message from your otter" notification" on my WATCH.
   (b) I opened the app on my WATCH and saw it.
   (c) I opened the app on my PHONE and saw it.
   (d) Other: (please describe)

2. What were you doing when you noticed this animation?

3. What message you were trying to convey with this animation to your partner?

4. What kind of reaction to this animation were you expecting from your partner, if any?

Our records indicate that your partner reacted to the state otter above with this react otter.
<GIF of the react animation received>

1. What did you think of this react otter, if you recall seeing it?

2. Did your partner react to your state otter above in any other way through the app or outside of the app (e.g., through other apps, in-person conversation)? If yes, please explain.

3. Were there any state otters you saw, either in notifications or in the app, but did not send? If yes, why did you decide not to send those state otters?

**Receiving your Partner's State Otter**
Our records indicate that you received the following state otter animation from your partner today around (time). Please answer the questions below about viewing your partner's state otter animation.

<GIF of the state animation received>

1. What did you think of your partner's state otter animation when you received it?

2. What message do you think your partner was trying to convey with this animation?

Our records indicate that you reacted with the following animation:
<GIF of the react animation sent>

1. Why did you react in this way?

2. What message were you trying to convey with this reaction?

3. Were there any other ways you wish you could have reacted to your partner's state otter (e.g., other animations or messaging options)?

4. Did you react to your partner's state at all outside of app (e.g., through other apps, in-person conversation, etc.)? If yes, please explain.

# Mid-Study Interview Questions

**General**
1. What has your experience with the app been like overall so far?

**Phone vs Watch**
1. Did you find yourself using the app more on your phone or your watch? Why?
2. How did you feel about using one vs the other?
3. Were there any differences in how you used it on the watch vs the phone?

**Notifications**
1. You mentioned in some of the daily surveys that you've noticed notifications such as (describe notifications they mentioned).
    a. What did you think of these notifications?
    b. When did you decide to open them, vs dismiss or ignore them?
2. Did you use the quick react or share button features? When and why?
3. Did you ever look at the app on your own, without first seeing a notification?
    a. Was this on your phone or watch? Why?

**Sharing Behavior**
Now we're going to talk about the state otters you sent to your partner, that they may or may not have reacted to.
1. Here's an example of a state otter you sent to your partner on (time). They reacted with (describe any react otters).
    a. Can you talk about what was happening when you sent it?
    b. What did you think of your partner's reaction?
2. Here are some examples of the state otters you've sent over the past two weeks (show slide of top 5), and some examples of the react otters your partner sent to you. Note that these are not necessarily matched up with each other.
    a. Do you recall sending these state otters?
    b. Can you give me a few examples of those times?
    c. How did your partner react?
    d. What did you think about the way they reacted?
3. Were there any state otter animations, provided in the app or not, that you wish you could have sent to your partner? Why or why not?

**React Behavior**
Now we're going to talk about the otters you received from your partner, and may or may not have reacted to.
1. Here's an example of a state otter you received from your partner on (time).
    a. Can you talk about what you thought when you received this otter from your partner?

      b. Why did you decide to send that react otter? (or not react)
2. Here are some examples of the state otters you received from your partner over the past two weeks (show slide of top 5), and some examples of react otters you sent to them. Note that these are not necessarily matched with each other.
    a. Do you recall seeing these state otters from your partner?
    b. Can you give me a few examples of those times?
    c. What do you think they were trying to convey to you?
    d. How did you react, if at all?
    e. Why did you react in this way?
    f. (if they sent a react otter) How did you decide which otter to react with?
    g. What were you trying to convey to your partner?
3. In the examples you described, were there any ways you wish you could have reacted to your partner, but weren't able to within the app?
4. Were there any otter animations, provided in the app or not, that you wish you could have used to react to your partner? Why or why not?

**Communication Behavior**
Now I want to ask you about your communication more generally.

1. Please describe any instances in which you and your partner talked about the app.
2. How has using the app been similar or different to other ways in which you communicate with your partner?
3. Do you find you're communicating similar or different things with the app versus other ways you communicate?
4. Did you ever use the app in conjunction with other communication channels, such as in conversation or with any of the apps you just mentioned? For example, sending an otter and then mentioning it in another app, or your partner receiving your otter and then commenting on it in another app.
5. (if mentioned) You mentioned in your daily surveys that your partner reacted to your otter outside of the app as well, for example, (describe their survey response). Could you describe this in more detail?
6. Have there been any changes to the way you communicate with your partner since the start of the study?

# Exit Interview Questions
**General**
1. What has your experience with the second version been like overall?
2. What did you think about the app update after the mid-study interview?
3. What was your understanding of what the update was?
4. Did you have any expectations for this update?

5. What did you think about the heart rate sensing?

6. Did you notice any differences in the way the app worked?
7. Did you notice any differences in the way you used this version, compared to the first version?
8. Which version do you prefer? Why?
9. Aside from the app changing, was there anything different about your circumstances from the first half of the study (e.g., your location, your job, ways you and your partner communicated)?
10. There are lots of reasons why participation in the study may have changed since the update. For example, life circumstances, or general decline in the novelty or motivation to use the app. Did you notice any differences in your participation in the study, including your use of the app, as the study went on?

**Notifications**
1. What did you think about the notifications in the second half of the study?
2. How did they compare to the notifications in the first half?

**Sharing Behavior**
1. How did knowing (if they knew) that the app sensed your state affect the way you sent your otter to your partner, if at all?
2. You mentioned for V1, that you sent your state otter when (describe instances they sent them from the mid-study interview). How did this compare to the times you sent your state otter with V2?
3. You mentioned for V1, that you sent your state otter because (describe reasons they sent them from the mid-study interview) How did this compare to the reasons you sent your state otter with V2?
4. In our last interview, you talked about how you sometimes expected certain reactions from your partner, both within and outside of the app. How did your expectations for your partner's reactions compare when you sent your sensed state otters in V2, if you had any?
5. Were there any times you saw a state otter that did match your mood that you decided not to send?

**React Behavior**
1. What did you think about seeing your partner's sensed state otter in V2?
2. How did knowing (if they knew) that you were viewing your partner's sensed state affect the way you reacted to your partner, if at all?
3. You mentioned for V1, that you sent a react otter when (describe instances/rea- sons for sending them described in mid-study interview). How did this compare to the times you reacted with react otters with V2?
4. You mentioned for V1, that you sent react otters that (describe the way they matched react otters in the mid-study interview). How did this compare to the way you reacted with react otters with V2?

5. You mentioned for V1, that you reacted on (describe other platforms they used to react stated in the mid-study interview). How did this compare to the times you reacted to your partner in the second half of the study?

**Communication Behavior**
1. Did you and your partner talk about the new version of the app at all? How did you talk about it? What did you talk about?
2. You mentioned in our last interview that V1 of the app was (describe what they said about how it compared to the other ways they communicate in the mid-study interview). How did V2 compare to the way you communicated with V1?
3. Have there been any changes to the way you communicate with your partner in the last two weeks?

**INTRO**

## Introduction

You will rate the characteristics of 17 different animations for a new messaging app for couples. This survey should take about 15-20 minutes.

**INSTRUCTIONS**

## Background

We are creating a new messaging app for couples, in which users can send animations as standalone messages to each other. Here are some example messages a user, represented by a light brown otter, might send:

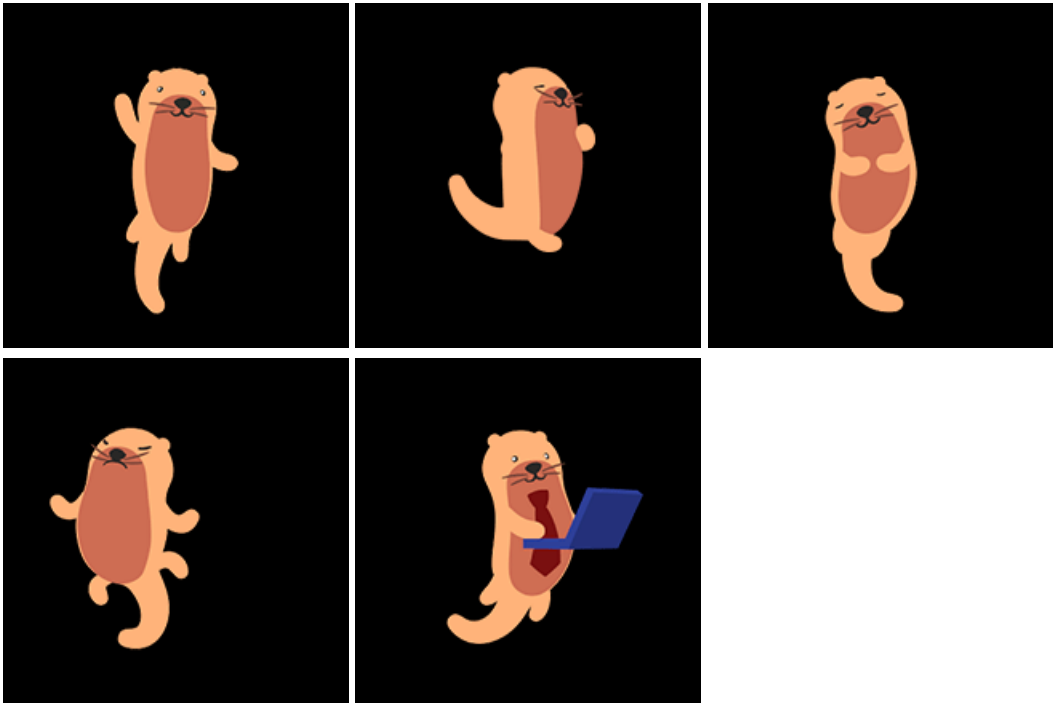

## Background

Users can respond to messages they receive from their partner using **reaction animations**. For example, Person A (represented by the light brown otter) may send the following animation to their partner, Person B:

Person A's message

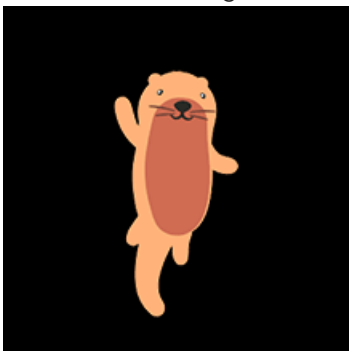

Person B (represented by the dark brown otter) may respond with the following **reaction animation**:

Person B's reaction

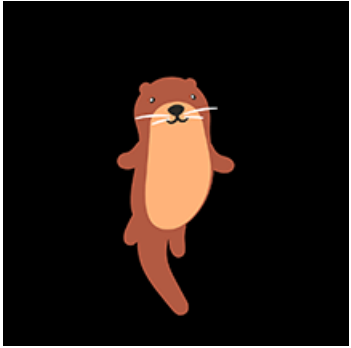

## Instructions

For the following questions, you will rate 17 **reaction animations**. Imagine using the animations you see as a standalone **reaction** to an animated message that a close partner sent (e.g., significant other, best friend, family member), as described on the previous page. In some animations, your partner will be represented by a light brown otter. Please answer the questions based on your first instinct.

**Neutral**

Answer the following questions for the **reaction animation**, sent in response to a close partner, below**:**

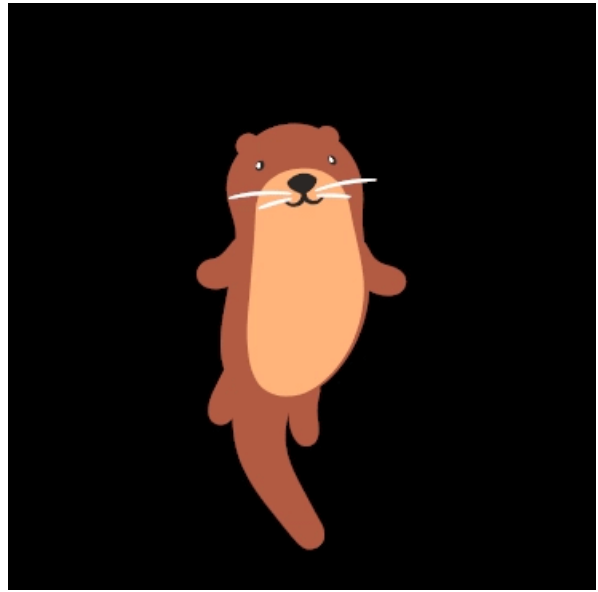

To what extent do you believe this animation expresses a *negative* versus *positive* reaction?

Negative  ○ ○ ○ ○ ○  Positive

To what extent do you believe this animation expresses a *low energy* versus *high energy* reaction?

Low energy  ○ ○ ○ ○ ○  High energy

To what extent do you believe this animation demonstrates *closeness (i.e., intimacy and familiarity) towards your partner*?

Not at all    ◯ ◯ ◯ ◯ ◯    Very much

To what extent do you believe this animation shows **caring and concern towards your partner**?

Not at all    ◯ ◯ ◯ ◯ ◯    Very much

To what extent do you believe this animation demonstrates **understanding of your partner's feelings**?

Not at all    ◯ ◯ ◯ ◯ ◯    Very much

To what extent do you believe this animation requests **follow-up action (after you've sent the animation) from your partner**?

Not at all    ◯ ◯ ◯ ◯ ◯    Very much

To what extent do you believe this animation shows that **your partner is valued**?

Not at all    ◯ ◯ ◯ ◯ ◯    Very much

Please write one possible message that this animation conveys in words.

Imagine a message you received **from your partner** that would prompt you to respond with this animation. Please describe that message.

### Hand Holding

Answer the following questions for the **reaction animation**, sent in response to a close partner, below**:**

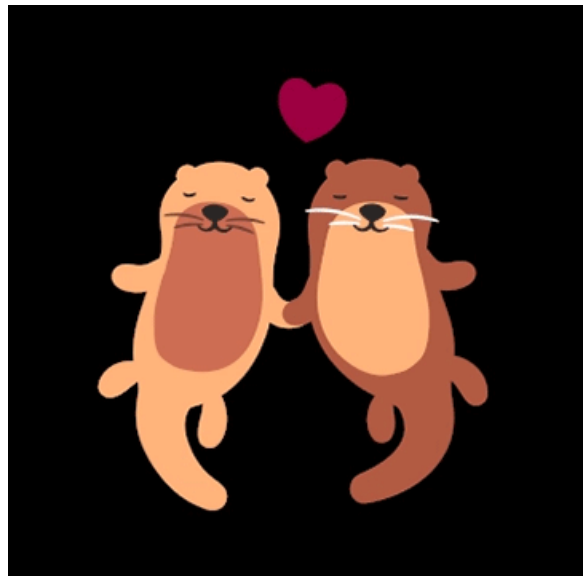

To what extent do you believe this animation expresses a **negative** versus **positive** reaction?

Negative  ◯ ◯ ◯ ◯ ◯  Positive

To what extent do you believe this animation expresses a **low energy** versus **high energy** reaction?

Low energy  ◯ ◯ ◯ ◯ ◯  High energy

To what extent do you believe this animation demonstrates **closeness (i.e., intimacy and familiarity) towards your partner**?

Not at all  ◯ ◯ ◯ ◯ ◯  Very much

To what extent do you believe this animation shows **caring and concern towards your partner**?

Not at all  ◯ ◯ ◯ ◯ ◯  Very much

To what extent do you believe this animation demonstrates **understanding of your partner's feelings**?

Not at all   ○ ○ ○ ○ ○   Very much

To what extent do you believe this animation requests **follow-up action (after you've sent the animation) from your partner**?

Not at all   ○ ○ ○ ○ ○   Very much

To what extent do you believe this animation shows that **your partner is valued**?

Not at all   ○ ○ ○ ○ ○   Very much

Please write one possible message that this animation conveys in words.

[                                                                                              ]

Imagine a message you received **from your partner** that would prompt you to respond with this animation. Please describe that message.

[                                                                                              ]



\end{document}